\documentclass[11pt,a4paper]{emulateapj}
\usepackage{epsfig}
\usepackage{amsmath}
\usepackage{natbib}
\usepackage{graphicx}
\usepackage[caption=false]{subfig}
\usepackage{wrapfig}
\usepackage[T1]{fontenc}
\usepackage{enumitem}
\submitted{Accepted by ApJ January 13$^{th}$, 2014}

\begin{document}

\title{Synthetic Observations of the Evolution of Starless Cores in a Molecular Cloud Simulation: Comparisons with JCMT Data and Predictions for ALMA}

\author{Steve Mairs}
\affil{Department of Physics \& Astronomy, University of Victoria, Victoria, BC, V8P 1A1, Canada: smairs@uvic.ca,}
\affil{National Research Council Canada, Herzberg Institute of Astrophysics, 5071 West Saanich Rd, Victoria, BC, V9E 2E7, Canada}

\author{Doug Johnstone}
\affil{Joint Astronomy Centre, 660 North A'ohoku Place, University Park, Hilo, HI 96720, USA}
\affil{National Research Council Canada, Herzberg Institute of Astrophysics, 5071 West Saanich Rd, Victoria, BC, V9E 2E7, Canada,}
\affil{Department of Physics \& Astronomy, University of Victoria, Victoria, BC, V8P 1A1, Canada}

\author{Stella S. R. Offner}
\affil{Department of Astronomy, 260 Whitney Ave, Yale University, New Haven, CT 06511, USA}

\and

\author{Scott Schnee}
\affil{National Radio Astronomy Observatory, 520 Edgemont Road, Charlottesville, VA 22903, USA}


\begin{abstract}

Interpreting the nature of starless cores has been a prominent goal in star formation for many years. In order to characterise the evolutionary stages of these objects, we perform synthetic observations of a numerical simulation of a turbulent molecular cloud. We find that nearly all cores that we detect are associated with filaments and eventually form protostars. We conclude that observed starless cores which appear Jeans unstable are only marginally larger than their respective Jeans masses (within a factor of 3). We note single dish observations such as those performed with the JCMT appear to miss significant core structure on small scales due to beam averaging. Finally, we predict that interferometric observations with ALMA Cycle 1 will resolve the important small scale structure, which has so far been missed by mm-wavelength observations.

\end{abstract}

\section{Introduction}
\label{introductionsec}

The coldest and densest regions of the interstellar medium are the places in which dust and gas form stars (see, for example, \citealt{benson1989}). Although the time a star spends on the main sequence as well as its subsequent evolution is one of the most well-understood problems in astrophysics today, the same statement cannot be made of the dense, dusty ``cores'' which are the progenitors of these objects. The opaque envelopes that have not yet formed protostars are of particular interest in that they reside at the intersection between the properties of the surrounding molecular cloud and its nascent suns.

A core, as defined in this work, is an object which has a relatively small mass and will form at most a few stars. Distinguishing ``starless'' cores from ``protostellar'' cores is the first step to answering several open questions in the field of star formation. For example, many starless cores have been measured to have masses several times that of their ``Jeans mass'' \citep{sadavoy2010}, the limit at which thermal pressure alone can provide adequate support against the self gravity of the object \citep{jeans1902}. One possibility is that there are poorly understood non-thermal support mechanisms, such as magnetic fields or turbulence, preventing the collapse of these cores. Another likely possibility, however, is that the cores have been misclassified as starless when in fact they are collapsing and dim protostars lie hidden within their dusty envelopes. In fact, there is strong evidence from recent interferometric observations (\citealt{howstarlessschnee,enochhydrostatic,dunhamsma,pineda2011}) that many cores classified as ``starless'' actually harbour Very Low Luminosity Objects (VeLLOs, objects with luminosities $\lesssim$ 0.1 $L_{\rm \odot}$; see \citealt{young2004,kauffmannvello,vellopaper,dunham2008}). 

Another avenue of study which relies on an accurate classification of starless and protostellar cores is the effort to link the prestellar Core Mass Function (CMF) to the stellar Initial Mass Function (IMF) (for example, \citealt{nutter2007}; \citealt{enoch2008}; \citealt{konyves2010}; \citealt{alves07}). It appears that more massive cores are typically found to harbour protostars (see, for example, \citealt{sadavoy2010big} and \citealt{raganhighmass}). As a result, if protostellar cores have been misclassified as starless and if the misclassification is more likely for particular masses, then any attempt to compare the observed starless CMF with the IMF is problematic. 

Finally, as \citet{kirk2005} describe, the lifetime of the subset of starless cores that will go on to form stars can be determined by comparing the relative number of starless cores and protostellar cores. Similarly, the number of protostars relative to more evolved young stars is proportional to the protostellar lifetime \citep{evans2009}. 

The most common method to classify cores as starless or protostellar is to identify cores via their dust continuum emission by using catalogues such as the ``Sub-millimetre Common-User Bolometer Array (SCUBA) Legacy Survey'' \citep{scubalegacy} and then attempt to identify embedded sources using infrared data (e.g. \citealt{jorgensen2007}; \citealt{sadavoy2010big}) such as the ``Molecular Cores to Planet Forming Disks Catalogue'' (c2d; \citealt{evans2003}, \citealt{evans2009}). Due to the high optical depth of these dusty envelopes, however, extinction can obscure and even completely hide dim protostars in the centre of these structures leading to errors in the core classification. In an attempt to explore the veracity of the non detections of embedded infrared objects, recent studies have utilised interferometric spectroscopy of a variety of protostellar and outflow tracers such as CO, SiO (2-1), HCO$^{+}$, and N$_{\rm 2}$H$^{+}$ \citep{pineda2011,howstarlessschnee,lackofsubstructureschnee}.
 
In order to gain further insight into the misclassification of starless cores while investigating the effectiveness of observational techniques, we compare starless and protostellar cores observed in the controlled environment of a simulated turbulent molecular cloud. We analyse the formation and evolving properties of dense structures in the same manner as real observations taken with SCUBA at the JCMT with the added benefit of knowing precise locations and masses of forming protostars. This analysis concentrates on the observed stability of a given object, as defined by the Jeans mass, near the time in which collapsing regions begin to form.

This paper is organised as follows: Section \ref{datasec} describes the numerical simulation. Section \ref{methodssec} outlines the methods we use to simulate observations and describes our method to identify objects and derive their stability. We present the bulk properties of the observed objects and their general evolution including stabilities, densities,  and protostar/envelope relationships as ``observed'' by SCUBA in section \ref{timesec}. Section \ref{ALMAsec} gives the results of the simulated interferometric observations. In section \ref{discussionsec}, we discuss the results. Section \ref{conclusionsec} presents concluding remarks.   

\section{Simulations}
\label{datasec}

In this paper, we analyse a series of snapshots from a hydrodynamic simulation of a turbulent molecular cloud that is forming stars. This simulation was previously presented in \citet{Offner13} (simulation Rm6), in which it was used to study the chemical distribution in molecular clouds. We briefly summarise the numerical procedure and parameters below.

The simulation was performed with the {\sc orion} adaptive mesh refinement (AMR) code \citep{truelove98,klein99}, and it includes large-scale driven turbulence, self-gravity, and sink particles \citep{Krumholz04}.   The simulation was first driven for two crossing times without gravity and then evolved for a global free-fall time with gravity.  Sink particles were inserted at the finest AMR level when the local density violated the Truelove criterion for $J=0.25$ \citep{truelove97}. This corresponds to a  mass density of 4.6 x 10$^{-16}$ g cm$^{-3}$ ($n_{\rm H} = 1.2 \mathrm{\:x\:} 10^{8}$ cm$^{-3}$). 
Throughout this work, the terms ``sink particle'' and ``protostar'' will be used interchangeably when discussing the simulation. The basegrid is $256^3$ cells and the run has 4 AMR levels.

The bulk properties of the simulation were chosen to represent a typical Galactic low-mass, star-forming region. The simulation domain has a length of 2 pc and contains $\sim$600 $M_{\rm \odot}$, which corresponds to an average number density of $n_{\rm H} = 1300$ cm$^{-3}$. The Mach number, $\mathcal{M}_{\rm 3D}$=6.6, was set so that the simulated cloud is approximately virialised and satisfies the linewidth-size relation (e.g., \citealt{MandO07}).

The simulation was run for one global free-fall time of 0.95 Myr. At the final time, the cloud contains 88 protostars and has a star formation rate per free-fall time of 0.18. Since the simulation does not include magnetic fields or stellar feedback, the sink particles represent an upper limit on the true star formation \citep{Offner09c,commercon11,hansen12}. 
Since the collapse has not been followed down to the sizes of individual protostars, and feedback such as outflows is not included, the sink particles likely over-estimate the stellar mass by a factor of $\sim 3$ \citep{matzner00,enoch07,alves07}. The most massive sink particle formed throughout the simulation is 8.5 $M_{\rm \odot}$.
	
\section{Synthetic Observation Methods}
\label{methodssec}	

\subsection{Single Dish Synthetic Observations}

				The simulation was placed at a distance of 250 pc to represent the Perseus molecular cloud as this region has been well-studied by c2d \citep{evans2003,evans2009} and other surveys \citep{kirkperseus,hatchellperseus,sadavoy2010big}. At this distance, 2 pc corresponds to 1650\arcsec. We analyse column density maps integrated along each of the x, y, and z directions. Each integrated image was gridded to 512 x 512 square pixels 3.22\arcsec $\:$on a side. 
				
				For the optically thin case, as we have here, it is simple to convert from column density, N, to flux, $S_{\nu}$:\\ $S_{\nu} = N\kappa_{\nu}B_{\nu}$, where $B_{\nu}$ is the Planck function and $\kappa_{\nu}$ is the opacity calculated at frequency $\nu$. The flux can then be related to the core mass via Equation 2 in \cite{sadavoy2010} (modified for a typical core temperature of 10 K):		 
				 
\begin{equation*}
\frac{S_{\rm 345}}{\mathrm{Jy}}=0.48\left(\frac{M_{c}}{M_{\odot}}\right)\left(\frac{d}{250 \mathrm{\:pc}}\right)^{-2}\left(\frac{\kappa_{\rm 345}}{0.01\mathrm{\:cm}^{2} \mathrm{\:g}^{-1}}\right) 
\end{equation*}
\begin{equation}
\times \mathrm{\:} \left\{\frac{\mathrm{exp}\left[1.7\left(\frac{10 \mathrm{\:K}}{T_{d}}\right)\right] - 1}{\mathrm{exp}(1.7)-1}\right\}^{-1}.
\label{masseq}
\end{equation}
Here, $S_{\rm 345}$ represents the flux received at 345 GHz (850 $\mu$m), $M_{c}$ is the core mass, $d=250$ pc is the distance to the source, $\kappa_{\rm 345}$ is the opacity at 345 GHz (see below),
				and $T_{d}=10$ K is the isothermal dust temperature. 
				
				We adopt an opacity value appropriate for a dusty protostellar core at 230 GHz, $\kappa_{\rm 230} = 0.009$ cm$^{2}$g$^{-1}$  \citep{ossenkopf1994},  in accordance with previous observations (e.g. \citealt{howstarlessschnee}). This value assumes MRN grains 
				with a thin ice mantle for a core with a density of 10$^{6}$ cm$^{-3}$ which is typical in nearby star forming regions (see \citealt{dougoph}, \citealt{howstarlessschnee}, \citealt{sadavoy2010big}). 
				
				By assuming a spectral index, $\beta$, where $\kappa_{\nu} \propto \nu^\beta$, one can extrapolate to other frequency values. For $\beta = 2.0$, which we adopt here, $\kappa_{\rm 345}$ = 0.0202 cm$^{2}$g$^{-1}$.
				
				The flux maps are smoothed to 20\arcsec $\:$$\:$to compare against the smoothed SCUBA catalogues\footnote{The JCMT beam is 15\arcsec $\:$but the SCUBA observations to which we are comparing were smoothed to 20\arcsec $\:$\citep{scubalegacy}.}. Then, to further match the observations, we remove large-scale structure by smoothing the same images
				 by 120\arcsec $\:$$\:$and subtracting this smoothed map
				from the former 20\arcsec $\:$$\:$maps (see also \citealt{kirkperseus}). In total, we analyse 68 simulation outputs distributed between $t=0-1t_{\rm ff}$. For each of these outputs, we consider each projection separately. Note that the mass of each core changes depending on the projection (see Section \ref{clumpfind} below) but, if detected multiple times, each detection will be considered as an individual core. The sink particle masses remain unchanged over each projection.			
				
\subsection{Core Definition}					
\label{clumpfind}

                                 To extract the bulk properties of the core population and analyse their time evolution, we use the automated routine CLFIND2D \citep{williams1994}. 
                                 
                                 The lowest flux level which defines the boundary of an observed core is 0.09 Jy/beam set by comparison with the observations. This threshold is defined by the opacity and, assuming the material is isothermal, it is equivalent to the mass depth to which we are sensitive (see Equation \ref{masseq}). Thus, the choice in our opacity value sets our scaling between the observations and the simulation.

                                 Each core's radius is then compared to the full width at half maximum (FWHM) of the smoothed 850 $\mu$m SCUBA observations, 20\arcsec. If the core is smaller than this,
				it is deemed spurious and removed from the analysis (since it might be noise). For the non-spurious cores, we convert
				the measured flux into a mass by inverting Equation 1. 
				
				Once the mass is attained, an object's stability can be analysed using simple assumptions. We estimate the Jeans mass, $M_{\rm J,c}$, of an identified core by applying the simple scaling relation presented by \cite{sadavoy2010},

\begin{equation}
				M_{\rm J,c} = 1.9\left(\frac{T_{d}}{10\mathrm{\:K}}\right)\left(\frac{R_{\rm c}}{0.07\mathrm{\:pc}}\right)M_{\rm \odot}.
\label{jeansmasseq}
\end{equation}
				Here, $T_{d} = 10$ K is the (assumed) isothermal dust temperature and R$_{\rm c}$ is the radius of the core. 
								
				To determine the stability of each core, we compare the mass attained from inverting Equation \ref{masseq} to the Jeans mass calculated
				by Equation \ref{jeansmasseq}. If $M_{\rm c}$ $\geq$ $M_{\rm J,c}$, the object is defined as ``super-Jeans'' an unstable configuration which should show signs of gravitational collapse 
				if thermal pressure alone were counteracting the force of gravity. $M_{\rm c}$ $<$ $M_{\rm J,c}$ represents a stable, ``sub-Jeans'' object which would not be expected to collapse
				since the thermal pressure within the assumed spherical object would be more than enough to balance the gravitational forces. 
				
				Once we determined the stability parameter using only the envelope mass,  we correlated the positions of protostars (sink particles) with the CLFIND2D objects. If a protostar lies within 75$\%$ of the circular radius (see below) of the centre of a core, we define the core to be protostellar. Therefore, a comparison can be made, for example, between the cores that are observed to be stable (cores without protostars) and the cores with evidence of collapse (due to the fact that they have embedded protostars). We tracked the growth of protostellar and starless core masses through time along with density, stability, and position. 
				
				To visually display the cores on the flux maps, the images include circles and squares corresponding to the size of each core at 
				the location of each of the core centroids (see Figure \ref{simfig}). Circles represent cores that are Jeans unstable (M $\geq$ $M_{\rm J,c}$) 
				and squares denote cores with masses less than the calculated Jeans mass. Plus signs symbolise the location of protostars. We constructed movies\footnote{Movies and plots are available at http://www.astro.uvic.ca/$\sim$smairs/research/starless/movies/movies.html} by stitching together images of sequential outputs (labeled at the top of each frame, see Figure \ref{simfig}).		
				
				 Since protostellar masses cannot be directly measured, adding the protostellar mass to the core mass does not make observational sense but allows us to track the stability of an object. When more than one protostar is associated with a given core, the protostellar masses are simply added.
				Without sufficient feedback, however, the available material will continue to accrete onto a given protostar. Without outflows, protostellar masses may be overestimated by up to a factor of $\sim3$ \citep{matzner00,enoch07,alves07}.
				
				Once a protostar is formed the inner envelope will be heated. Observationally this makes it harder to convert from observed 850 micron flux to mass and thus makes Jeans stability investigations difficult for protostellar cores. For the simulations investigated here, however, the lack of included heating makes the mass determination and stability analysis straightforward.
				
				Note that the mass, size, and density scales of objects we extract are dependent on the large-scale structure in the image. To investigate, we performed structure identification for maps that had not undergone the 120\arcsec $\mathrm{\:}$ scale removal. In these cases we find large reservoirs of mass, which are strongly associated with filamentary structure (see Section \ref{filaments}), surrounding each core. These extended zones are approximately twice the size of cores identified when the large-scale structure is removed. It is useful to consider that it has been previously determined that CLFIND2D works reasonably well when a field is sparsely populated with discrete objects, but struggles to sensibly pick out important structures when the field is ``crowded'' (see, for example, \citealt{pinedaclfind}).

\subsection{Interferometric Synthetic Observations}				
\label{interferometermethods}

The Common Astronomy Software Applications (CASA)\footnote{see http://www.casa.nrao.edu} package is used to simulate 100 GHz Atacama Large Millimetre Array (ALMA) Cycle 1 interferometric observations of several individual cores  in order to compare with the single dish results. We choose a five second ``snapshot'' integration time to match real observations; 90 seconds in total is required to create a mosaic for each core.

 Beginning with the simulated column density maps, we again generate flux maps as described above. 
For a given core, we use a square 2 arcminutes on a side centred on the object's coordinates as the input sky model (the simulated flux map).  Then, we construct the UV visibilities for the most compact arrangement of antennas using the CASA package \textit{simobserve}. We employ \textit{simanalyze} to Fourier transform these visibilities into the image plane. 

In this study, the simulation has similar densities and temperatures to the Perseus cloud. Therefore, the declination of the observation is set to a reasonable approximation of the cloud's position: J2000 +30d00m00. 
We use a hexagonal stitching pattern for each mosaic.

All observations include thermal noise. 
In CASA, a robust atmospheric profile exists for the ALMA site including the altitude, ground pressure, relative humidity, sky brightness temperature, and receiver temperatures. In \textit{simobserve}, the user need only define the precipitable water vapour (pwv) which had a chosen value of 1.262 mm. With these assumptions, the noise for the ALMA Cycle 1 image is $\sim0.1$ mJy/beam. Since these simulated observations are only meant to note whether the source was detected, the maps are not ``cleaned''; we  simply convolve them with the point spread function of the instrument to produce a ``dirty'' map for analysis. This will not significantly alter the mass of the objects detected, however, a cleaned map would reveal more structure over a potentially larger area. 

Simulated observations using the Submillimetre Array (SMA) and the Combined Array for Research in Millimeter-wave Astronomy (CARMA) were also performed, but we did not detect any cores in the resulting maps.  

\section{Results from Single Dish ``Observations''}
\label{timesec}

				Over the course of one free-fall time, we traced several core properties both qualitatively and quantitatively. The following sections describe our main results.

\subsection{Filamentary Structure}				
\label{filaments}

The idea that filaments play a role in in the formation of dense cores has been entertained for several decades (see, for example, \citealt{schneiderfilaments}). The growth of instabilities in filaments and subsequent fragmentation have been numerically analysed in both the linear (e.g. \citealt{inutsuka1992}) and nonlinear regimes (e.g. \citealt{inutsuka1997}). Robust simulations of isothermal, self gravitating cylinders have revealed several properties of filaments such as characteristic temperatures, external pressures, densities, and radii which are consistent with Herschel observations \citep{fischera2012,fischeraprop2012}. The Herschel observations show that filaments are present throughout many star forming regions \citep{andrefilaments}. Additionally, \citet{hacar2013} performed observations of C$^{18}$O(1-0), N$_{\rm 2}$H$^{+}$(1-0), and SO($J_{N}=3_{\rm 2}-2_{\rm 1}$) in the Taurus star forming region with the 14 m FCRAO telescope (and supplemented the data with APEX 870 $\mu$m and IRAM 30 m 1200 $\mu$m) where they found that cores appear to form in a two step process. First, velocity-coherent filaments form. Then, these large structures fragment into cores. \citet{kirkfilamentflow} and \citet{friesenfilamentflow} have found evidence of filamentary accretion flows in the Serpens star forming region using a variety of spectral lines observed with the ATNF Mopra 22 m telescope and the K-Band Focal Plane Array at the Robert C. Byrd Green Bank Telescope, respectively.  \citet{hennemann2012} suggest that gravitationally unstable filaments are the driving factor for star formation in the Cygnus X region with a specific emphasis on the DR21 ridge. Furthermore, \cite{myersfilaments} shows that young, embedded star clusters are associated with multiple filaments. It has also been shown that filamentary geometry is ideal for the growth of small-scale perturbations that lead to large scale collapse in a preferential dimension (the length of the filament) and that filaments containing only a few Jeans masses can easily fragment (for a thorough description, see \citealt{pon2011}).   

Of course, some detections of filaments may be attributed to multiple beam diluted objects so careful investigation of properties such as velocity coherence from spectral fitting are necessary to characterise a structure with certainty. As suggested by the authors above and references therein, a typical filament within a molecular cloud has a width of $\sim$0.1 pc. We removed scales of $0.15\mathrm{\:pc}$ (120\arcsec $\:$at the distance of the Perseus star forming region) from the analysis of the simulation presented here, however, to focus on the cores themselves and not their parent structures.      

With this in mind, the convolved simulation hosts cores which are 
				forming almost exclusively along what appear to be dense filamentary structures. As the simulation evolves in time, the objects appear to travel
				along these striations, fragmenting and coalescing into a variety of morphologies (see movies).
				
\begin{figure*}
\centering
\subfloat{\label{}\includegraphics[width=8.5cm,height=8.5cm]{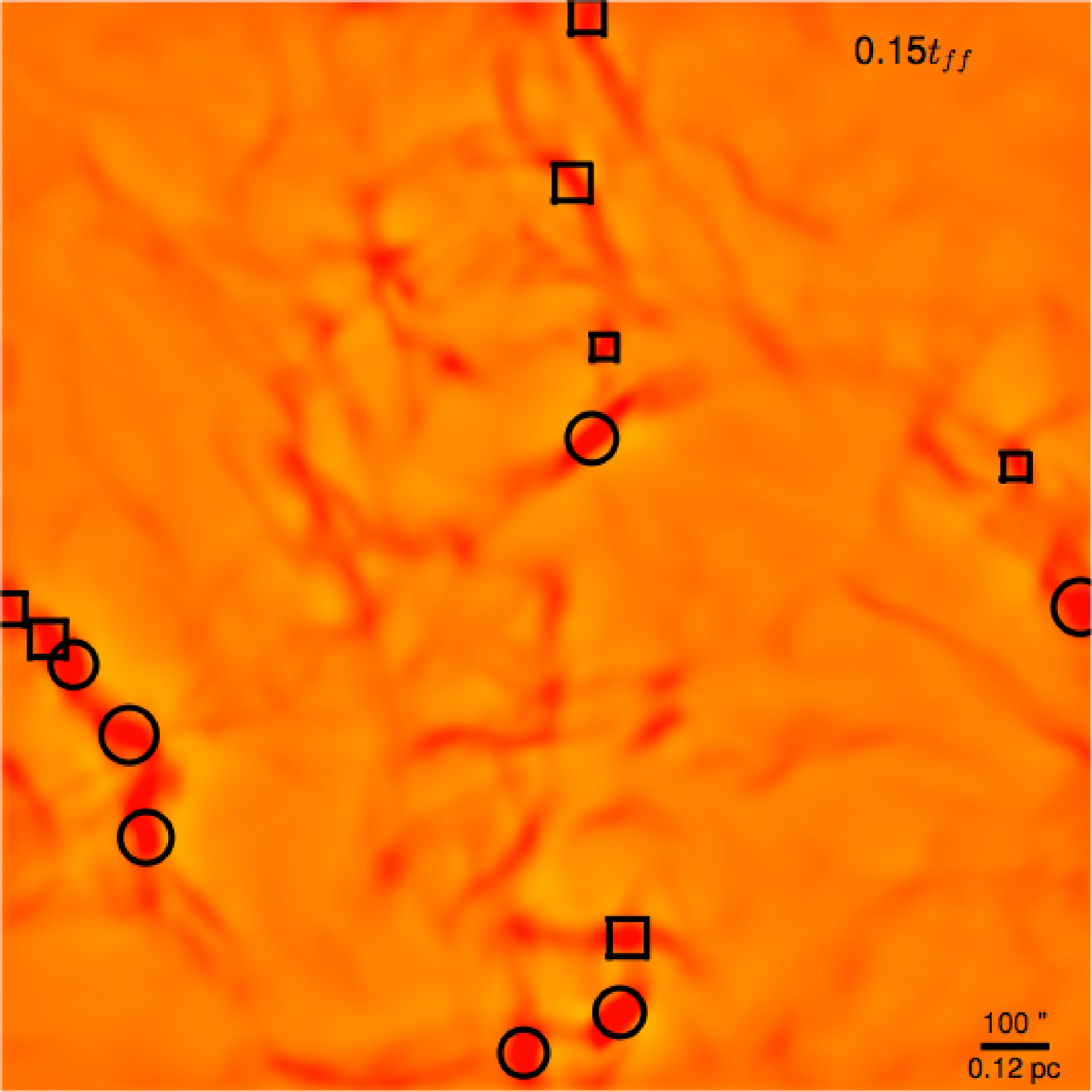}}
\subfloat{\label{}\includegraphics[width=8.5cm,height=8.5cm]{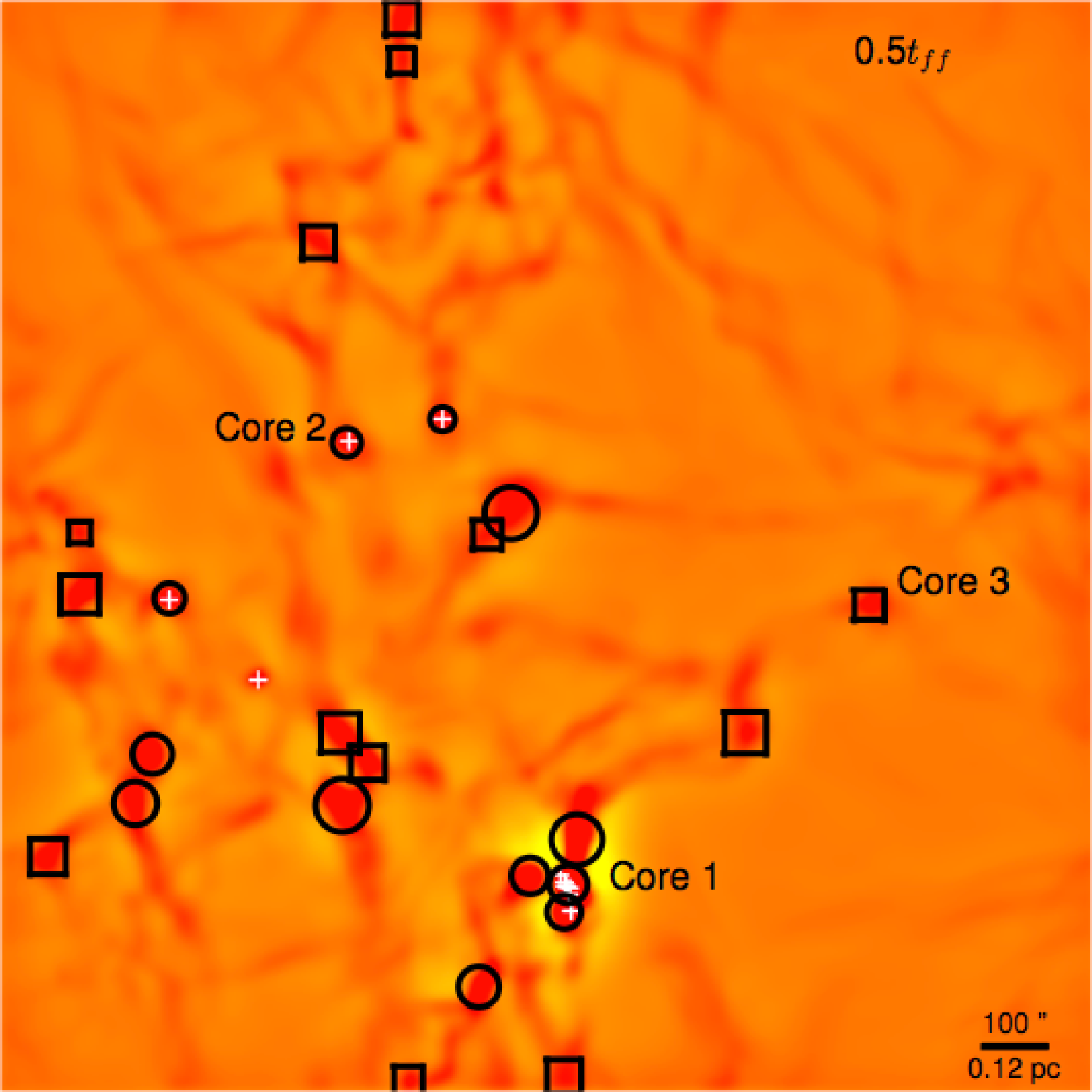}}\\
\subfloat{\label{}\includegraphics[width=8.5cm,height=8.5cm]{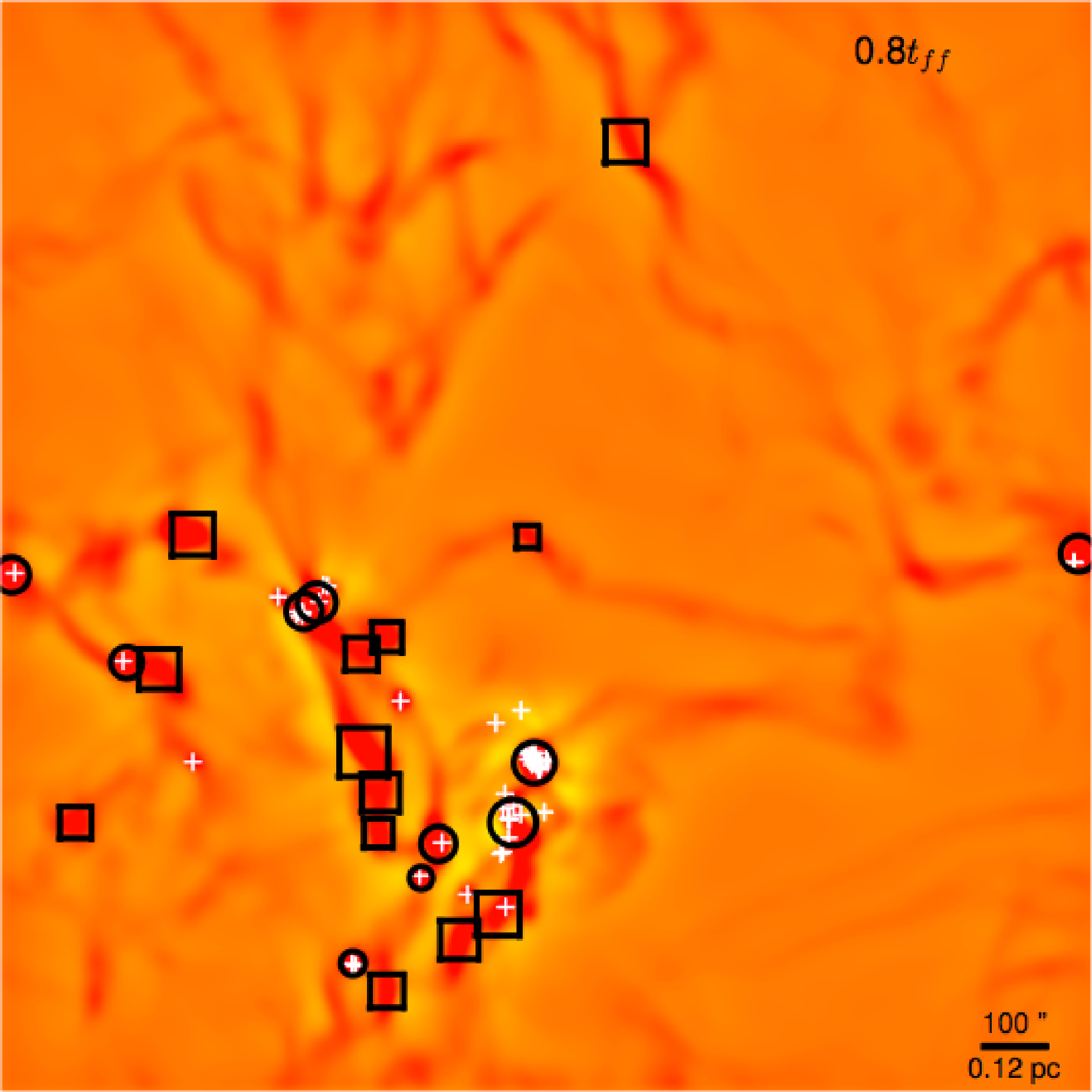}}
\subfloat{\label{}\includegraphics[width=8.5cm,height=8.5cm]{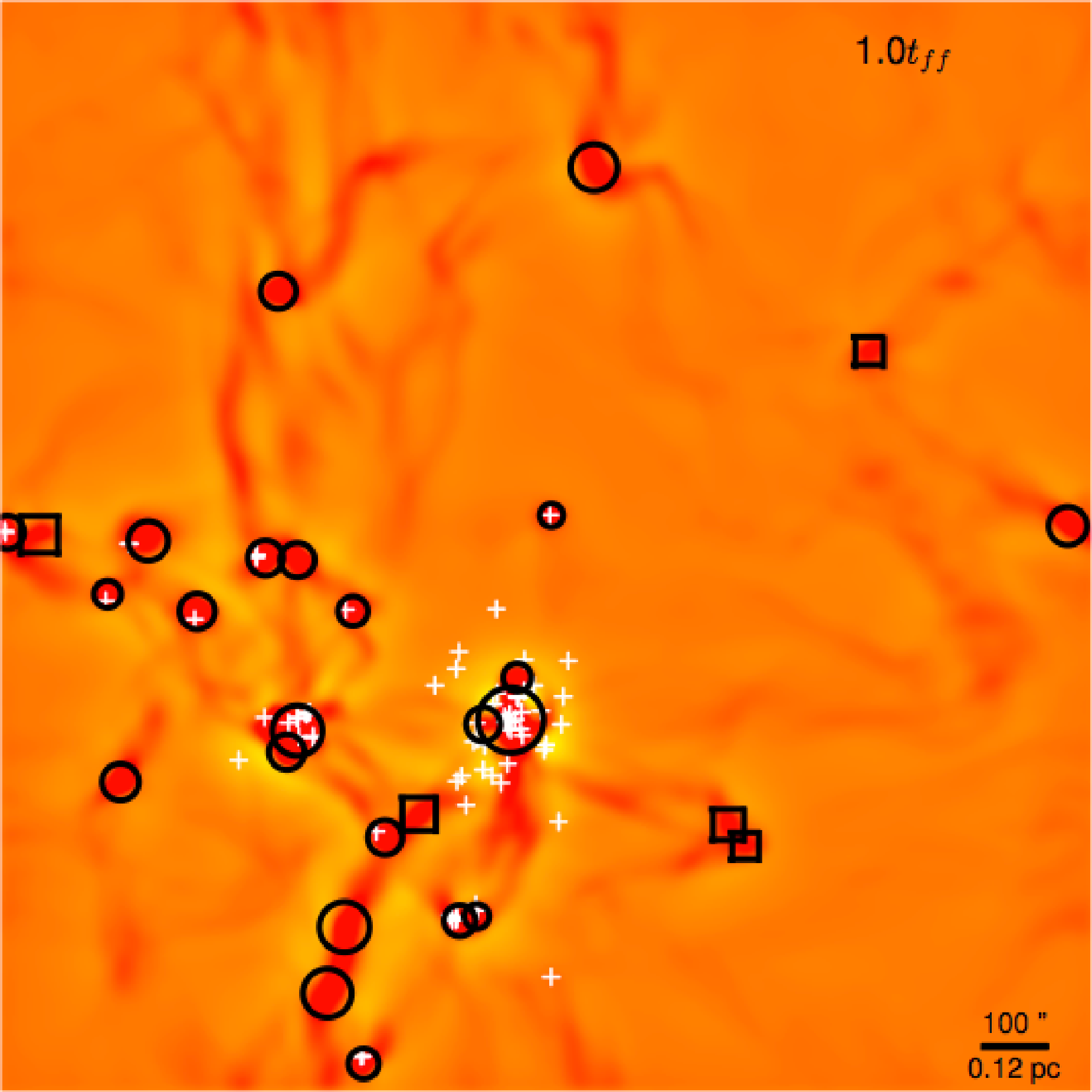}}\\
\caption{Four snapshots ranging from ``early'' times to ``late'' times (t = 0.15$t_{\rm ff}$ to t = $t_{\rm ff}$, see labels). Protostellar masses have been included in the stability calculations. The Y-dimension integrated images are shown. Circles represent unstable cores, squares show the locations of stable cores, and plus signs display the locations of protostar formation sites. Three cores that we study in greater detail are highlighted in the top right panel.}
\label{simfig}
\end{figure*}

				Figure \ref{simfig} shows images of four of the sixty-eight snapshots. The top left panel 
				represents an early time in which no protostars have formed; the bottom right panel represents one free-fall time, $t_{\rm ff}$. The filaments stand out clearly and their qualitative association with the identified cores is obvious.

\subsection{Bulk Properties of the Ensemble}
\label{bulk}

\begin{figure*}		
\centering
\subfloat{\label{}\includegraphics[width=8cm,height=7.6cm]{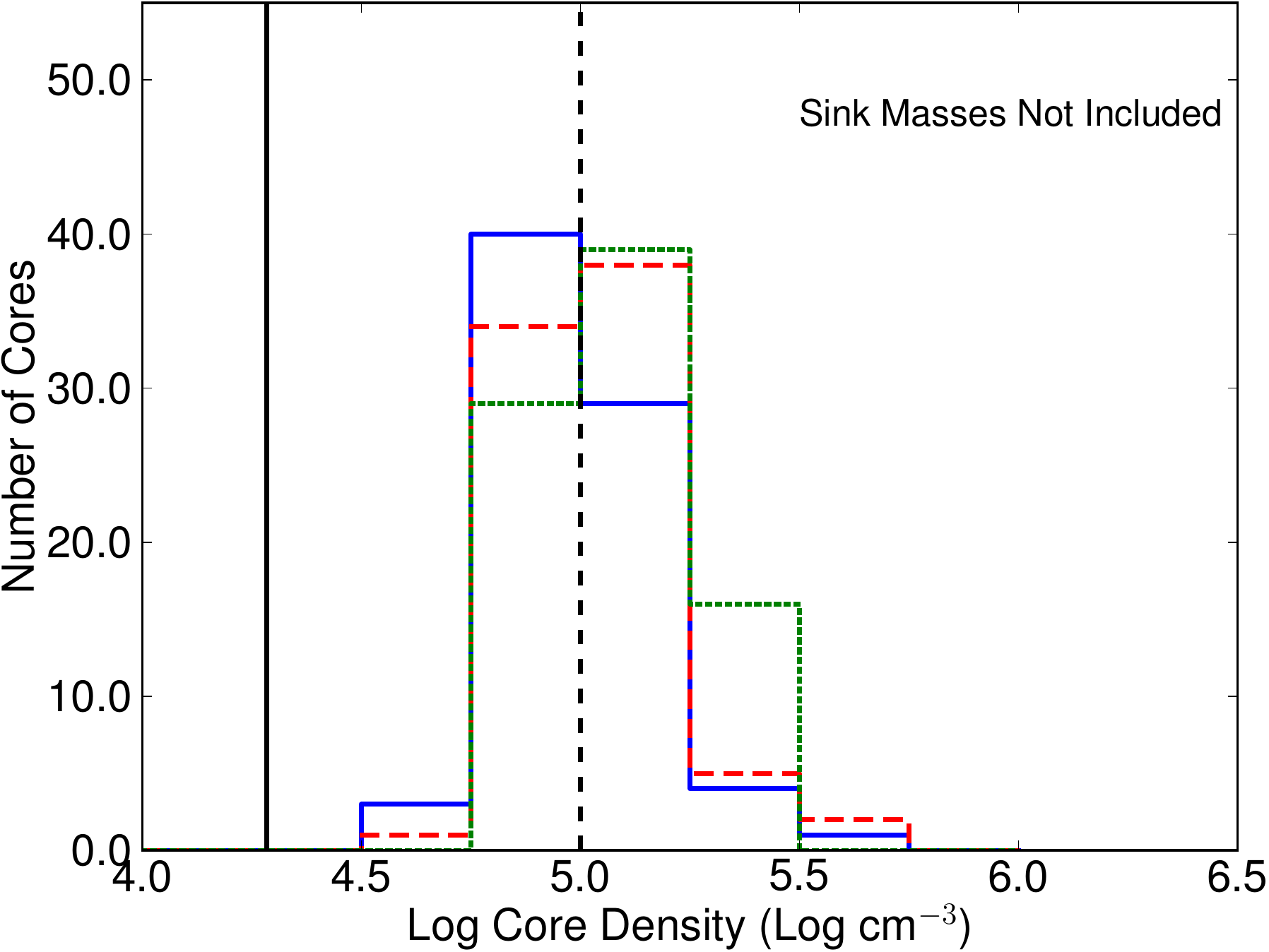}}
\subfloat{\label{}\includegraphics[width=8cm,height=7.6cm]{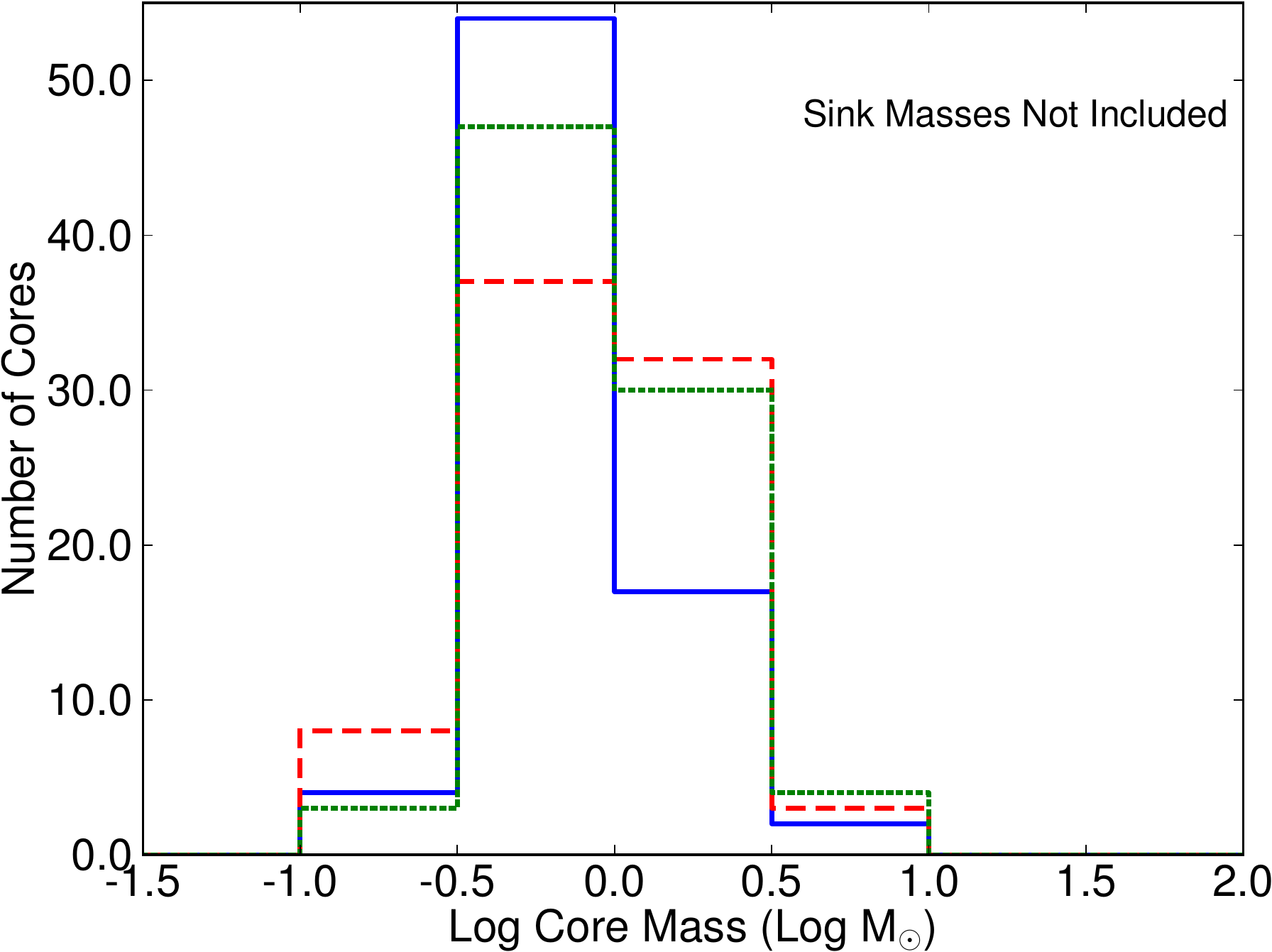}}\\
\subfloat{\label{}\includegraphics[width=8cm,height=7.6cm]{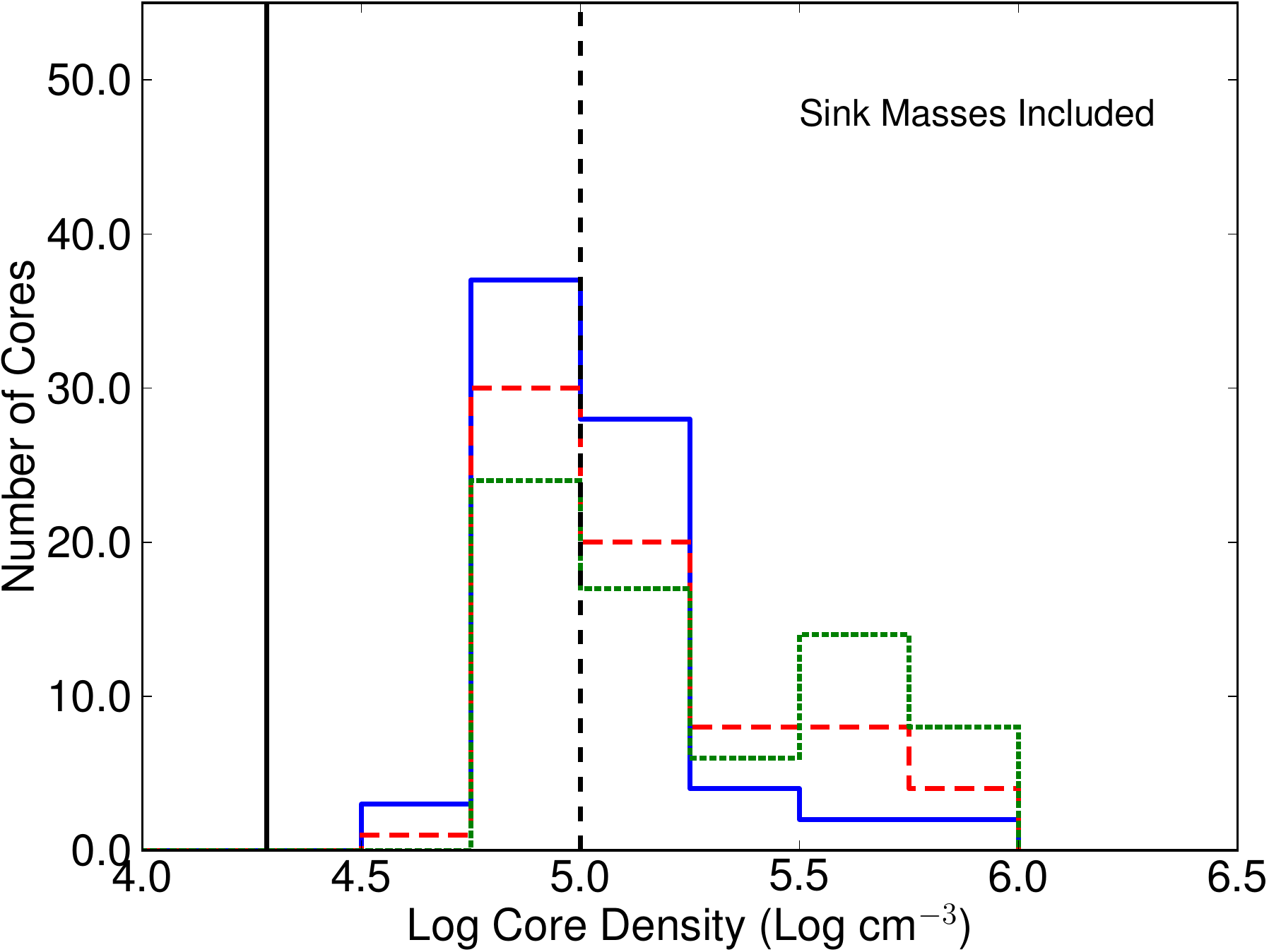}}
\subfloat{\label{}\includegraphics[width=8cm,height=7.6cm]{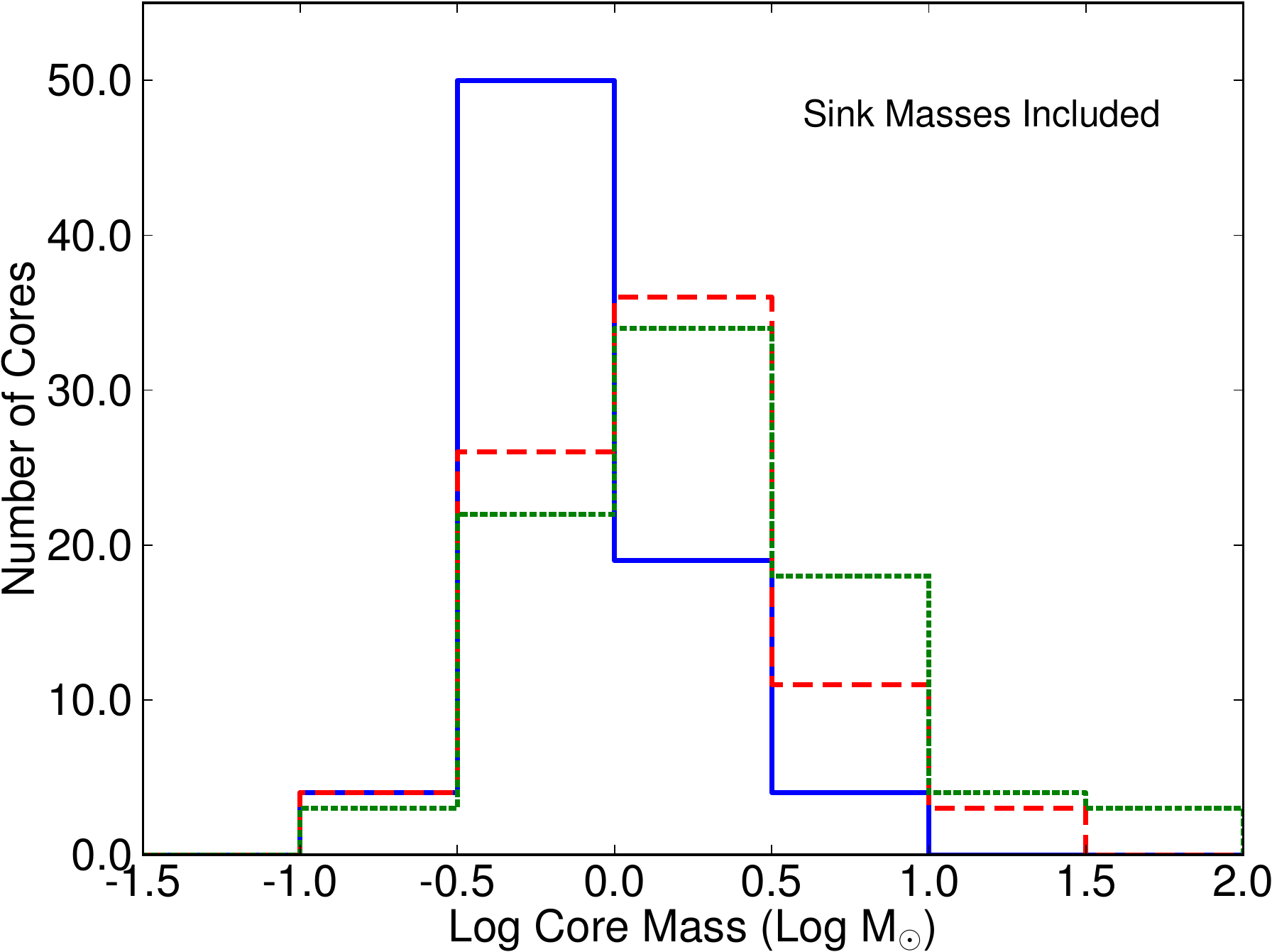}}\\
\caption{The core density (left column) and mass (right column) distributions at three different times: 0.5$t_{\rm ff}$ (solid line), 0.8$t_{\rm ff}$ (dashed line), and $t_{\rm ff}$ (dotted line) including all three projections. The protostellar masses are not included in the top row (to emulate real observations of cores with hidden protostars); they are included in the bottom row. The solid vertical line in the density plots shows an estimate of the typical density of shocked regions. The dashed vertical line highlights a peak in the density distribution, the ``modal density''.}
\label{hists}
\end{figure*}

In this section we present the mass and density distribution of the identified cores, where we analyse cores identified from three orthogonal views. Figure \ref{hists} shows the distributions of core masses and densities at different times throughout the simulation. We calculate the density of each core by assuming the mass is uniformly distributed over the object as if it were a sphere with a radius determined by its projected area.

The top row of Figure \ref{hists} presents the dataset in which no protostellar masses were included in the mass determination in order to compare directly with observations. We note that the range in core masses and densities found in the simulation are consistent with real observations of Perseus \citep{sadavoy2010big,enoch2008}. The bottom row illustrates the core mass distribution including the protostellar masses in order to analyse the ``true'' stability of a given core. As time progresses, the core masses increase as gas accretes and collapses to higher densities. This is more evident in the bottom row than the top. New cores are identified throughout the simulation and therefore an approximately constant low mass population of objects is present throughout the later stages of the simulation. 

At the end of the simulation when protostars are not taken into account in the mass estimate, the median core mass and the mean core mass are both found to be $0.89 \mathrm{\:M}_{\rm \odot}$. Including the mass of protostars yields nearly equivalent median and mean masses of approximately 1.9 $M_{\rm \odot}$. The median and mean number densities of the dataset without protostellar masses are both  $n_{\rm H} = 1.3 \mathrm{\:x\:} 10^{5}$ cm$^{-3}$. The dataset including protostellar masses has a median number density of $n_{\rm H} = 1.8 \mathrm{\:x\:} 10^{5}$ cm$^{-3}$ and a mean number density of $n_{\rm H} = 2.7 \mathrm{\:x\:} 10^{5}$ cm$^{-3}$.

The density of a core can be compared with three reference values: the average density of the box, $n_{\rm 0}$, the density of a typical shocked region, $n_{\rm s}$, and the ``modal density'', $n_{\rm c}$. 

$n_{\rm 0} = 1.3 \mathrm{\:x\:} 10^{3}$ cm$^{-3}$. This is an order of magnitude lower than the density of any identified object. To estimate the compressed density, we consider a 1D, isothermal shock at the typical density: $n_{\rm s} = \mathcal{M}_{\rm 1D}^{2}n_{\rm 0} = 1.9 \mathrm{\:x\:} 10^{4}$ cm$^{-3}$ (see section \ref{densities} for more details). This value is shown by the vertical line in Figure \ref{hists}, which is close to the lowest densities of the cores. 

The ``modal density'' is an empirically derived density noted by the vertical dashed line in Figure \ref{hists}. Whether or not the masses of the sink particles are taken into account in the analysis, there appears to be a peak in the density distribution at a number density of $n_{\rm H} \sim 1.0 \mathrm{\:x\:} 10^{5} \mathrm{\:cm}^{3}$. Note that the tail of the distribution in Figure \ref{hists} is quite different depending upon whether we include the protostellar masses in the estimates, especially at later times.

\subsection{Core Stability}
\label{corestability}

\begin{figure*}		
\centering
\subfloat{\label{}\includegraphics[width=8cm,height=7.6cm]{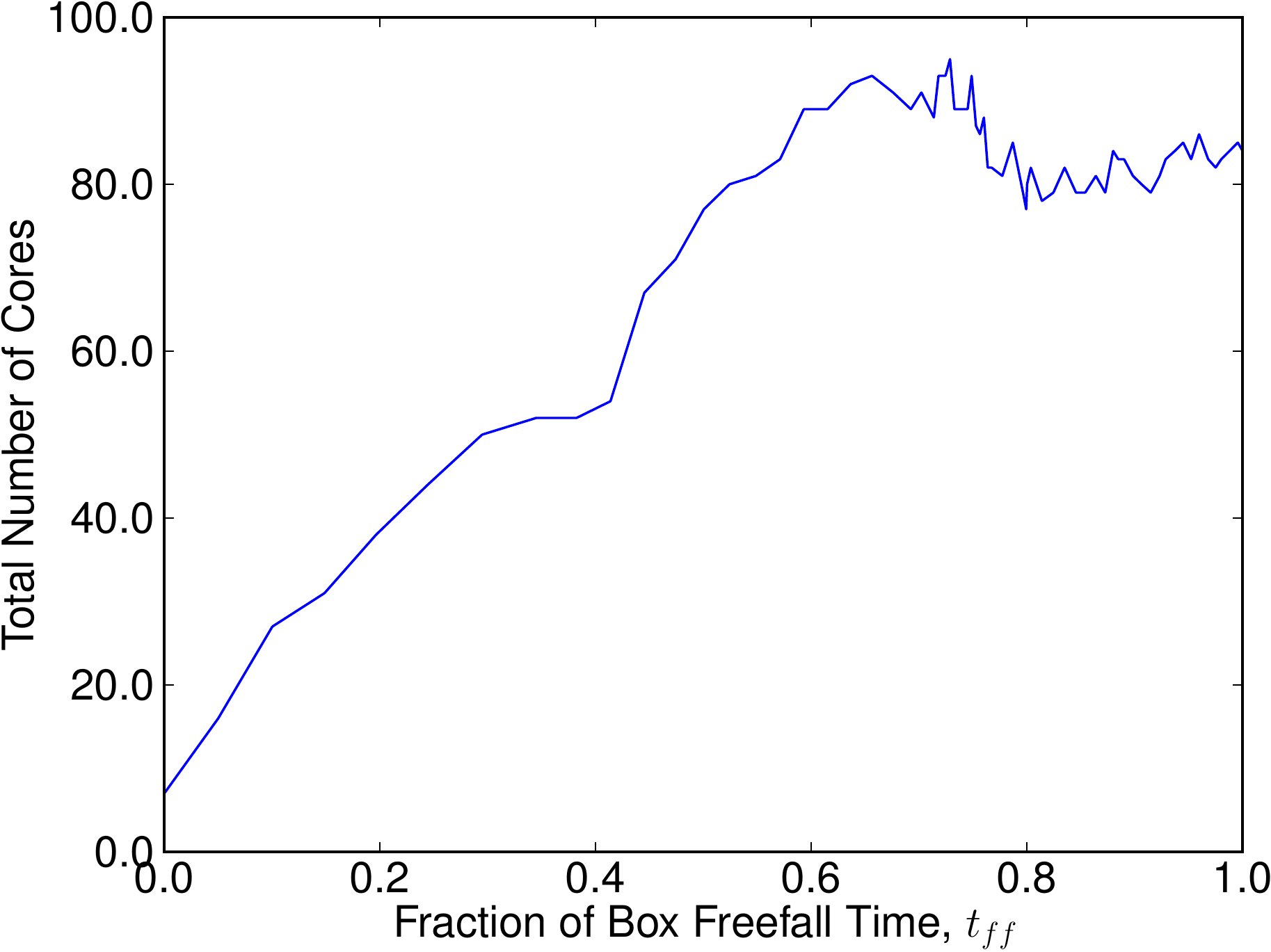}}
\subfloat{\label{}\includegraphics[width=8cm,height=7.6cm]{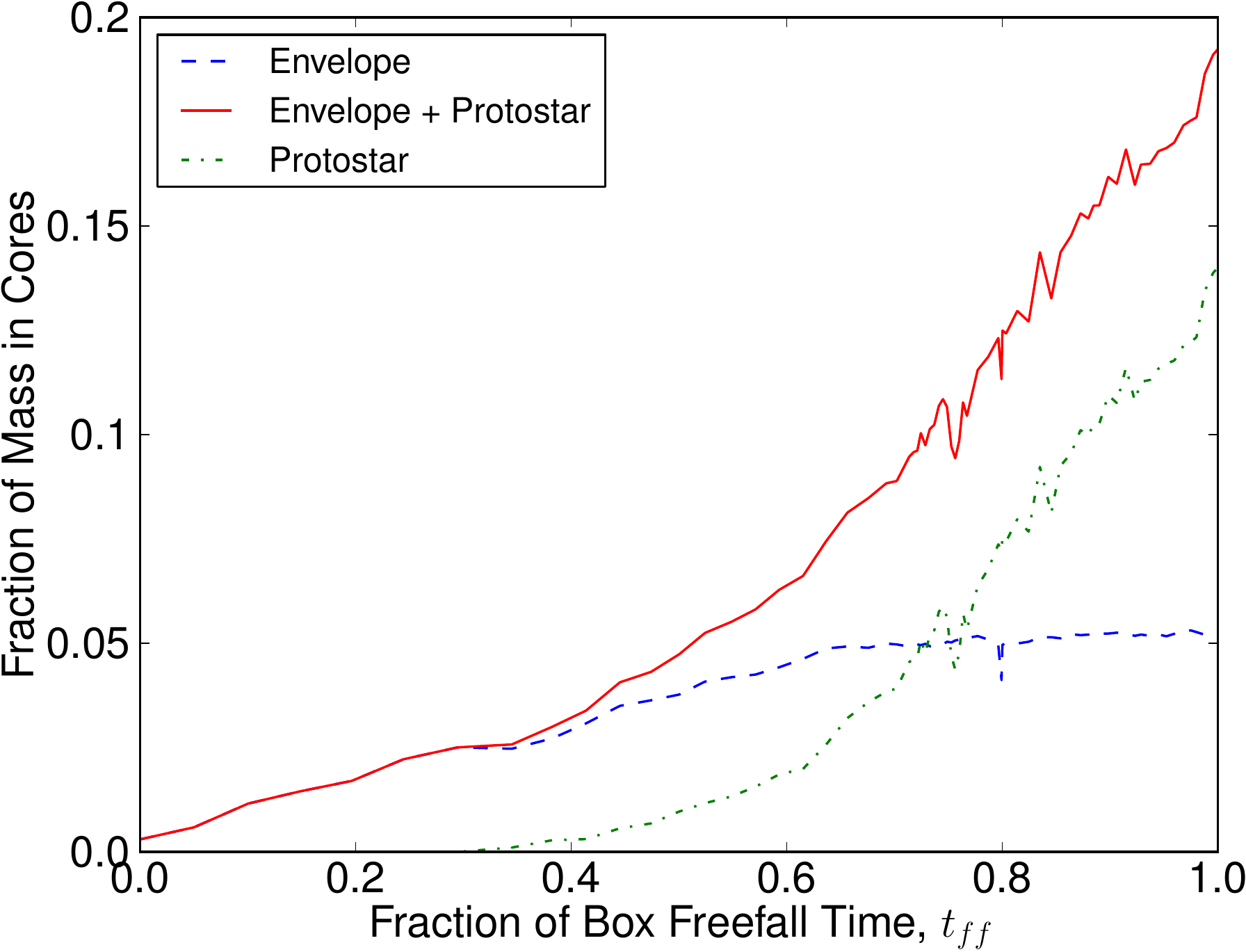}}\\
\caption{\textit{Left:} The total number of cores identified through time during the simulation. 
\textit{Right:} The fraction of the simulation's total mass contained within identified cores over all three projections. The dashed line represents the dust envelope mass only; the solid line shows the dust envelope mass as well as their contained protostar masses, the dash-dot line shows the protostellar structure mass only. Note that the large majority of the mass contained within cores is locked in protostars where it cannot be directly observed. After one free-fall time, protostellar masses account for 15\% of the 600 $M_{\rm \odot}$ box, or, 90 $M_{\rm \odot}$ while core envelopes account for $\sim$5\% of the mass of the box, or 30 $M_{\rm \odot}$.}
\label{totalnumbers}
\end{figure*}

				Over time, cores become visible because gravity produces densities which exceed the threshold of observability. Thus, more cores are identified as the simulation proceeds (see Figure \ref{totalnumbers}). At t = 0.5$t_{\rm ff}$ there are 78 identified cores over all three projections. 80 cores are identified by t = 0.8$t_{\rm ff}$, and the simulation ends with 86 identified cores at t = $t_{\rm ff}$.			
				
				As expected, the simulation begins
				with few identified cores, then mass accumulates and observationally stable starless cores begin to form in what appears to be a bottom-up fashion. As the simulation 
				proceeds, many of these cores become unstable and begin to form protostars (see right hand panel of Figure \ref{totalnumbers}). 
				
\begin{figure*}		
\centering
\subfloat{\label{}\includegraphics[width=8cm,height=7.6cm]{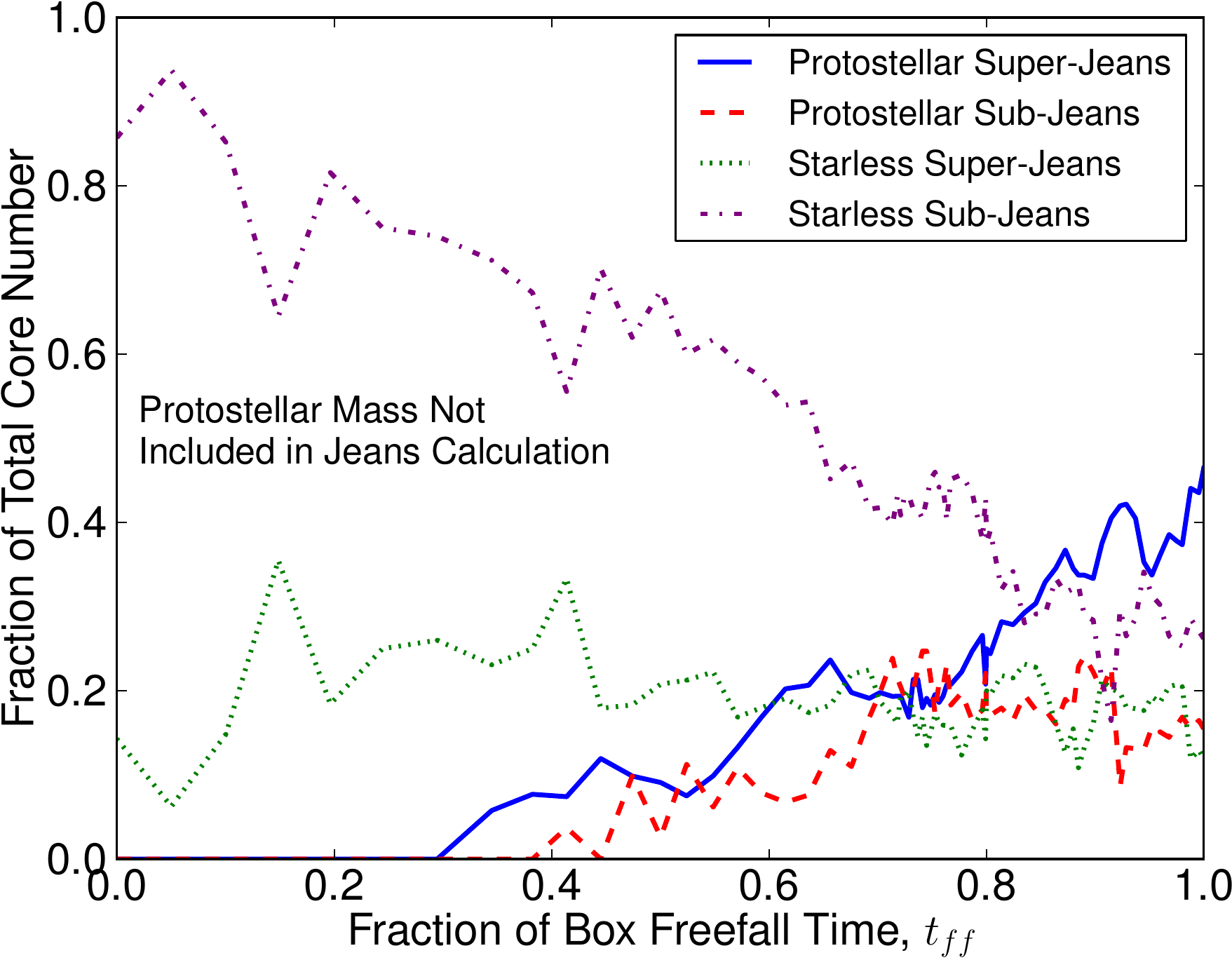}}
\subfloat{\label{}\includegraphics[width=8cm,height=7.6cm]{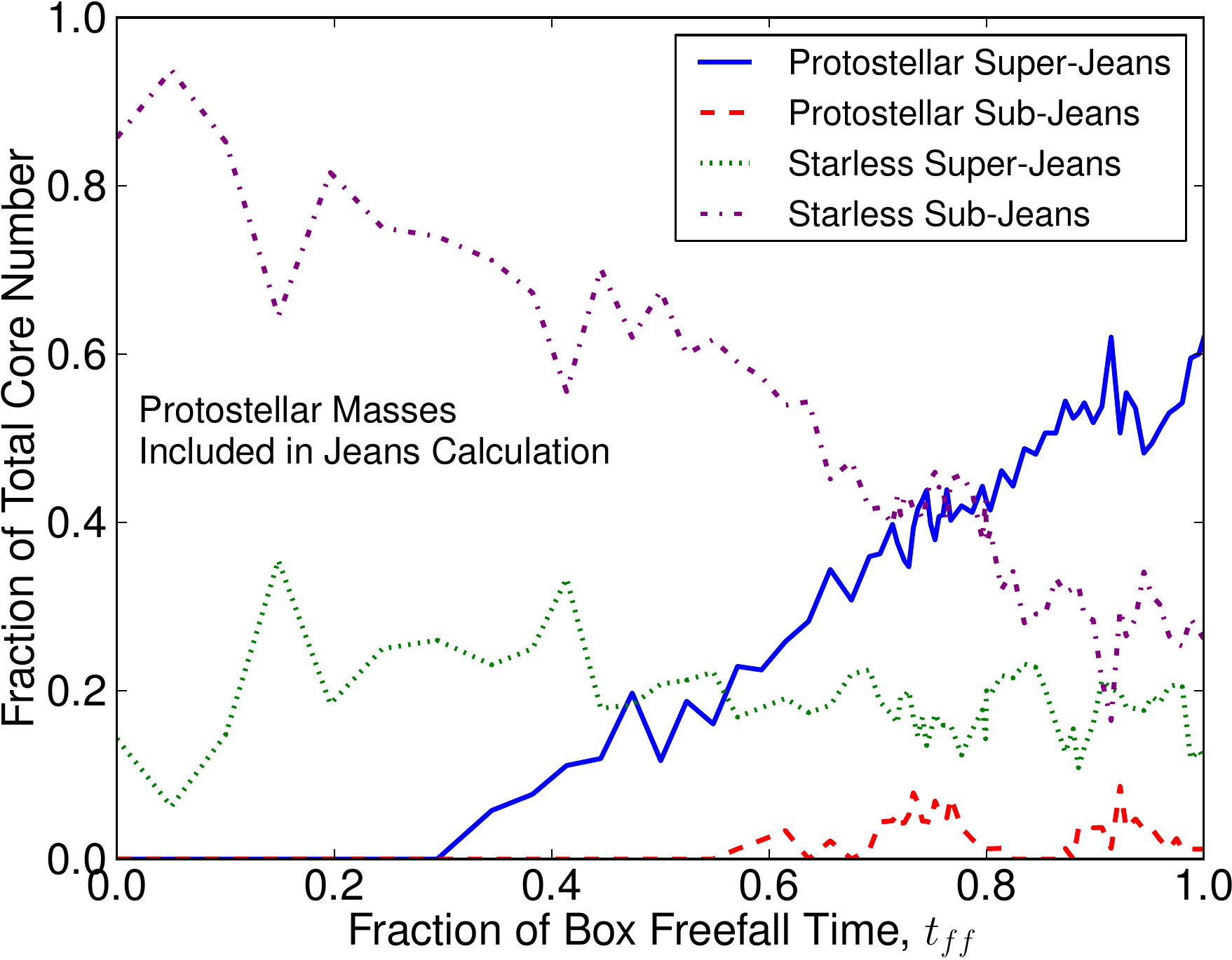}}\\
\caption{Different core stability states. \textit{Left:} Protostellar masses are not included in the analysis, \textit{Right:} Protostellar masses are included. $M_{p}$ is the protostellar mass.} 
\label{stats}
\end{figure*}

				 Figure \ref{stats} shows the number of cores in different stability states throughout the observed portion of the simulation for cases in which the dust envelope alone is taken into consideration (left panel) and when protostars are added to the flux maps (right panel). Together,
				the two panels show the evolution of cores defined as sub-Jeans (observationally stable), super-Jeans (observationally unstable to collapse), protostellar (contain a sink particle), and starless (contain no sink particles).

				It is important to note that when the protostellar masses are not taken into account, the true masses of the identified objects are underestimated and there is a population of cores that are deemed ``stable'' even though they are collapsing and forming protostars. This suggests that, when the embedded protostar is unobservable, the Jeans stability argument used by \cite{sadavoy2010} may not be sufficient to identify super-Jeans cores.
				
\begin{figure}		
\centering
\includegraphics[width=9cm,height=8cm]{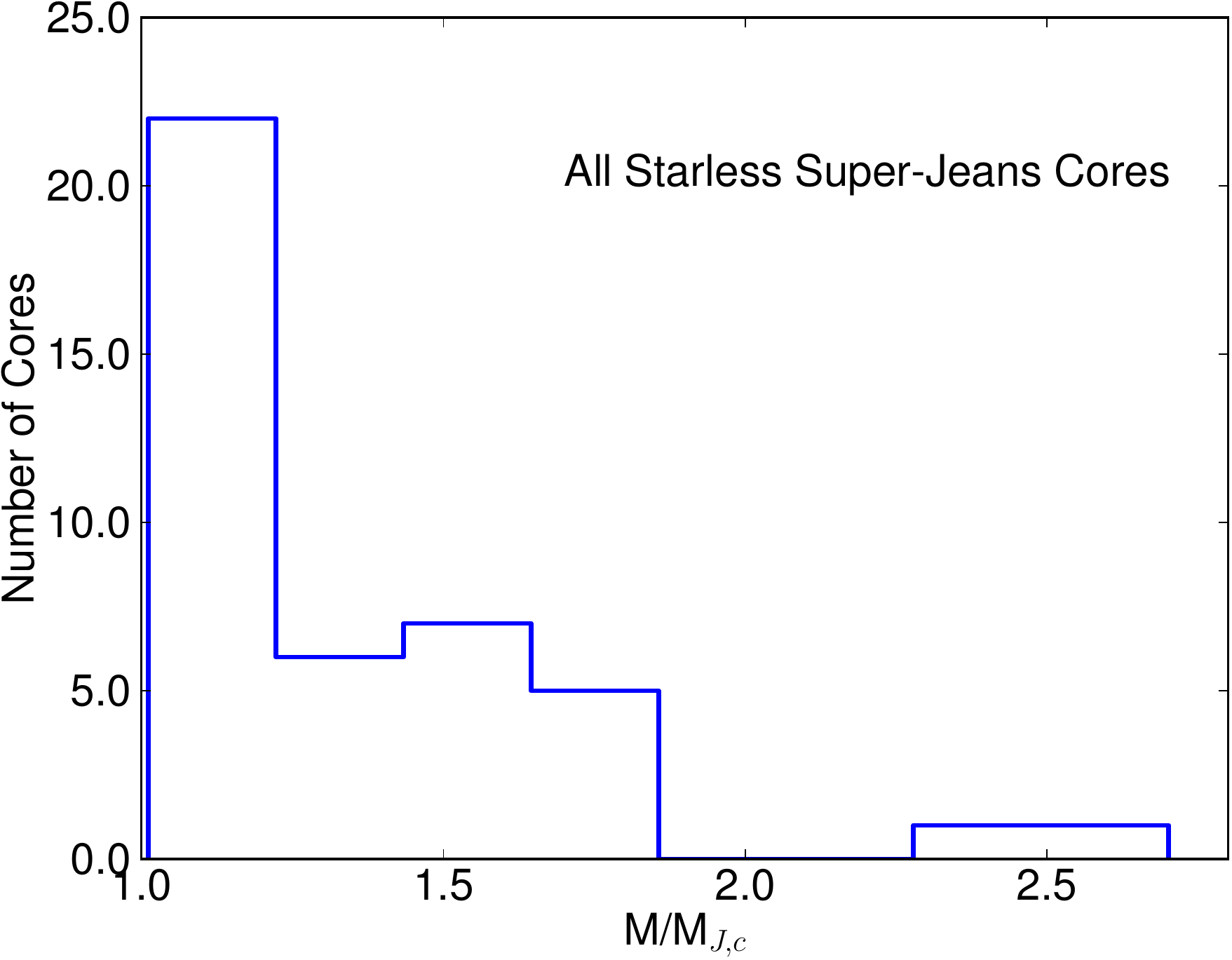}
\caption{The masses of the starless super-Jeans core population (the unstable prestellar cores) over all timesteps and projections in the simulation given in terms of their individual Jeans masses.}
\label{starlesscorepop}
\end{figure}

				Although we find super-Jeans cores without any protostars inside, it appears these objects are only marginally unstable. All but one of these cores have masses which are less than a factor of two greater than their respective Jeans masses (see Figure \ref{starlesscorepop}). As noted by \cite{sadavoy2010}, starless cores which satisfy  $M\geq$ $4.5M_{\rm J,c}$ are those which are deemed ``unusual relative to the protostellar cores''. We find no such cores throughout this simulation. 
					
				When protostellar masses are included, the observed core stability changes significantly. We see that practically all of the previously sub-Jeans cores which had protostars are now classified as super-Jeans as expected.
							
\subsection{Evolution}
\label{densitytrackssection}

\begin{figure*}		
\centering
\subfloat{\label{}\includegraphics[width=8cm,height=7.6cm]{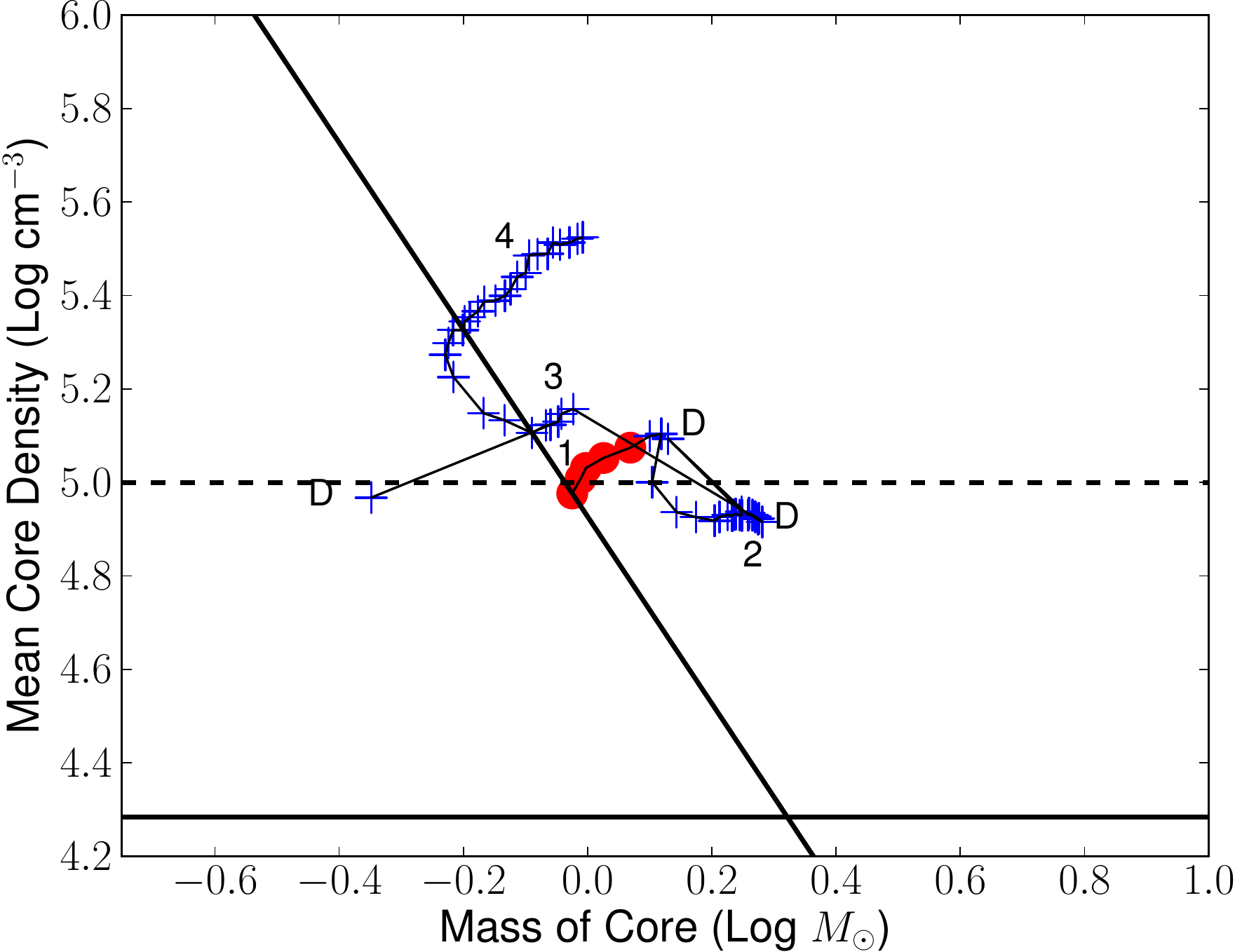}}
\subfloat{\label{}\includegraphics[width=8cm,height=7.6cm]{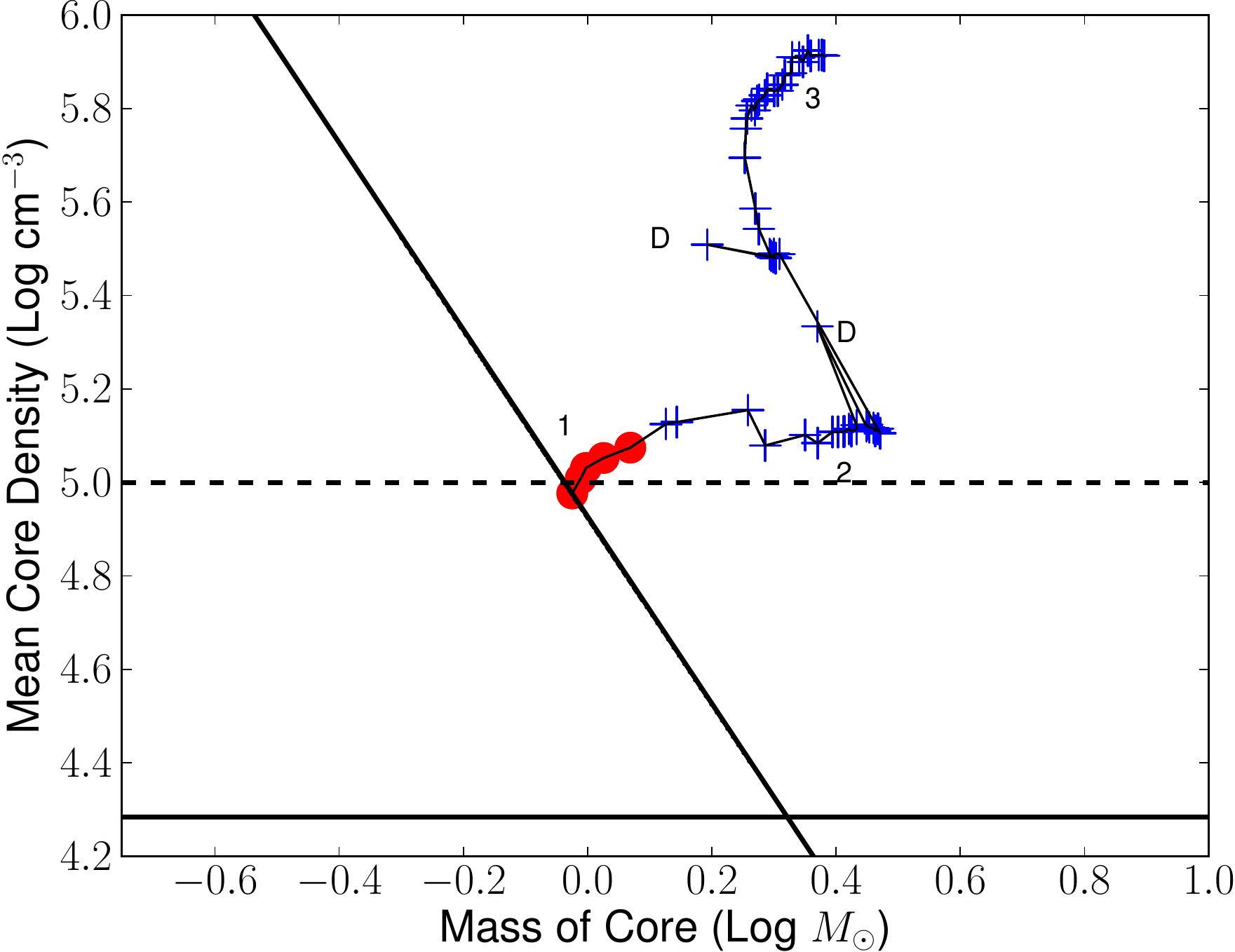}}\\
\subfloat{\label{}\includegraphics[width=8cm,height=7.6cm]{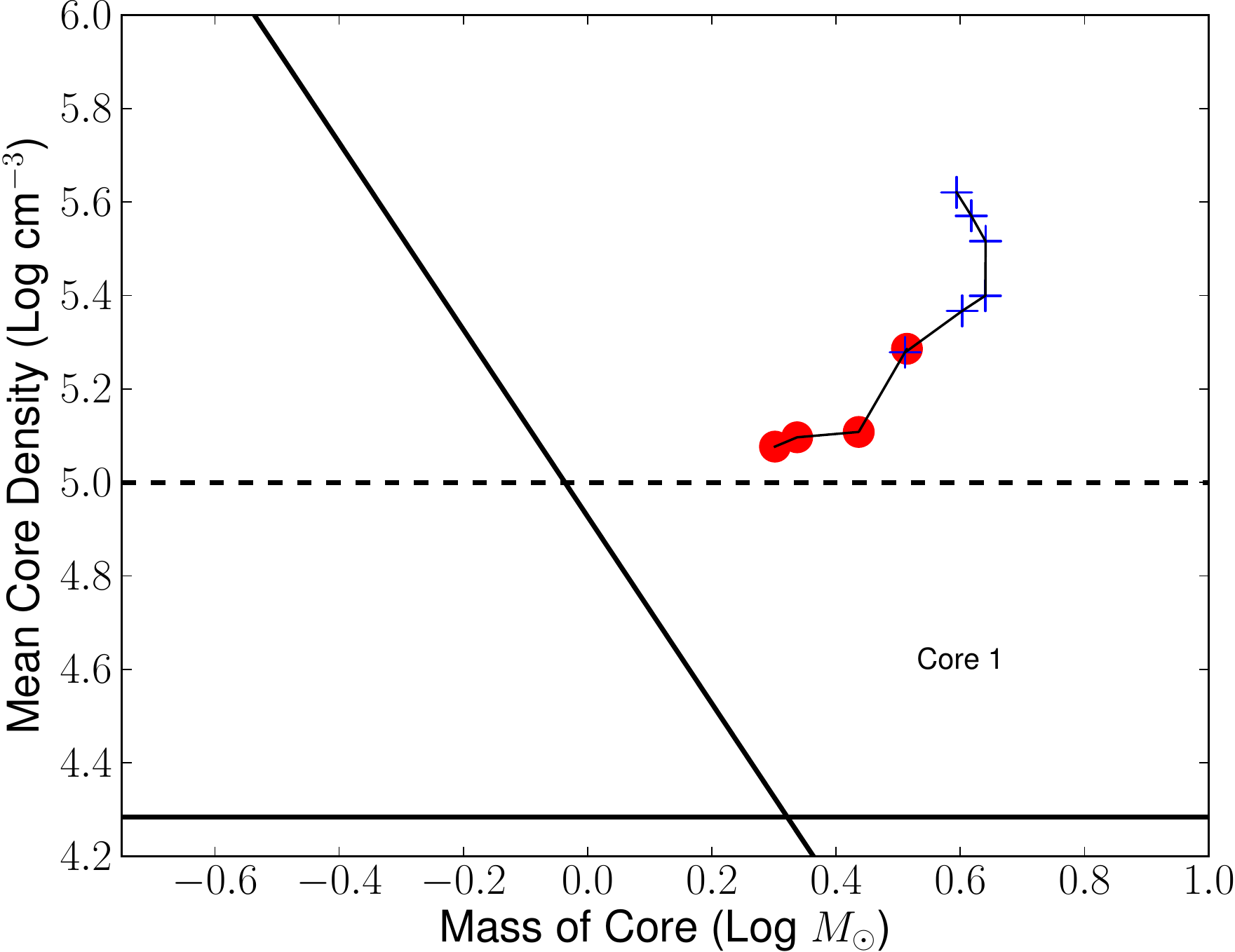}}
\subfloat{\label{}\includegraphics[width=8cm,height=7.6cm]{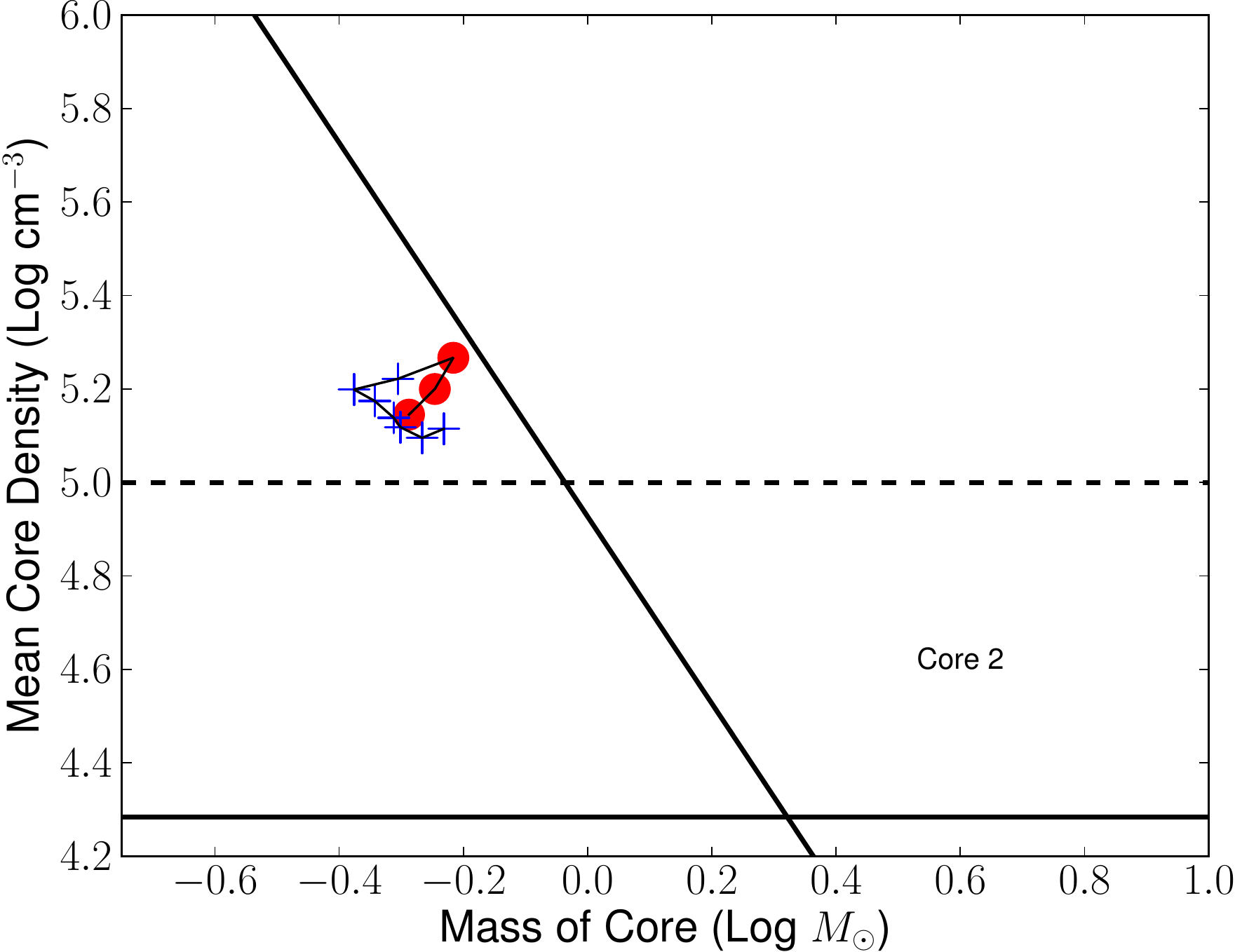}}
\caption{Top panels: One core tracked over all the outputs of the simulation.  \textit{Left:} Dataset in which protostellar masses are not included. \textit{Right:} Dataset in which protostellar masses are included. Bottom Panels:  Densities of two individual cores which form protostars tracked over a subset of the outputs of the simulation. The left panel shows the object labeled ``Core 1'' in Figure \ref{simfig} and the right panel shows ``Core 2''. Protostellar masses have not been taken into account for either of these latter cores. Circles indicate when the core does not contain a protostar within its boundaries. Plus signs indicate at least one protostar exists within the core boundaries. Points lying above the solid diagonal line are defined to be observationally unstable using Equations \ref{masseq} and \ref{jeansmasseq}; points lying below are classified as observationally stable. The solid horizontal line shows the fiducial shock density (see Section \ref{densities}). The dashed horizontal line shows the empirically derived ``modal density''. ``D'' represents a discontinuous feature introduced by CLFIND2D.}
\label{random}
\end{figure*}

				We selected several cores to study in greater detail. These cores are taken from isolated positions in the flux maps so other objects and protostars will not significantly affect the measurements (three are highlighted in Figure \ref{simfig}).
				We track one of these cores through the entire simulation and choose specific timesteps for the others based on the protostar formation time. Figure \ref{random} shows the mass and density of three example cores tracked over a number of times. 
								
				 In Figure \ref{random}, the solid horizontal line shows the density associated with the shocked material (see Section \ref{densities}). The dashed horizontal line highlights the ``modal density'' as defined in Section \ref{bulk}. The diagonal line indicates the minimum density for a core to be considered observationally unstable (Equation 2).

				It is evident that there are a few discontinuous jumps (annotated by ``D's'') in Figure \ref{random}. These are due to CLFIND2D itself and how it defines multiple objects. 
				
				Although a core is initially isolated, it fragments and coalesces with its
				pieces as it evolves. When an extended object becomes large enough to exceed the flux threshold between two regions, CLFIND2D draws a boundary
				between the regions and labels each as a separate object. Sometimes, this bifurcation lasts only for a brief period of time and the object reassembles into its previous configuration in the next timestep. Of course, a sudden decrease in radius becomes a sudden
				increase in density and vice versa. Both in the synthetic observation and in actual observations, there are occasionally multiple objects and filaments that have been ``smeared'' into one identified core due to the 20\arcsec $\:$smoothing.

				Both panels in Figure \ref{random} highlight specific regions of interest. In the left panel, which is directly comparable with real observations, stage 1 shows the beginning of the core evolution when a protostar has not yet formed. 
				
				There is a short period (stage 2) in which the mass increases but the density remains constant. A brief subdivision and merging takes place before the core splits into two distinct objects, losing mass and becoming far more dense as it enters stage 3. Another CLFIND2D identified bifurcation and amalgamation takes place when the defined flux threshold is briefly achieved before the object settles into its final evolutionary state (stage 4). From this point until the end of the simulation, the mass and the density both increase as in stage 1, but to a much larger degree. When considering the envelope mass alone, the evolution is less monotonic but still exhibits periods of collapse.

		The right panel of Figure 5 shows density versus mass including the protostellar mass. There are two obvious regions marked with ``D's'' in which CLFIND2D sub-divided, then merged this object. Beginning in stage 1, there is a very steady mass and density increase as this object forms a protostar and becomes observationally Jeans unstable shortly thereafter. After the early collapse, the density begins to level off for a short period of time. Meanwhile, the mass continues to increase, indicating that the radius must also be increasing before a large upward jump in density. Here again, the core bifurcates causing an abrupt density spike. During this period, the mass is essentially constant for the rest of the simulation, which suggests the core is simultaneously decreasing in size.

				The bottom left panel of Figure \ref{random} shows the evolution of a core which is recognisably Jeans unstable before a protostar forms.				The right panel shows an isolated core that we select at random in the simulation to highlight the diversity in the core population. When only the envelope masses are taken into account, each core resembles the left panel of Figure \ref{random}. When protostellar masses are included as shown there is a rapid evolution in the core properties. Both the mass and the density increase substantially over time, such that the cores appear observationally unstable.

\subsection{Protostar and Envelope Relationship}
\label{sinkenvelope}
				
\begin{figure}		
\centering
\includegraphics[width=9cm,height=8cm]{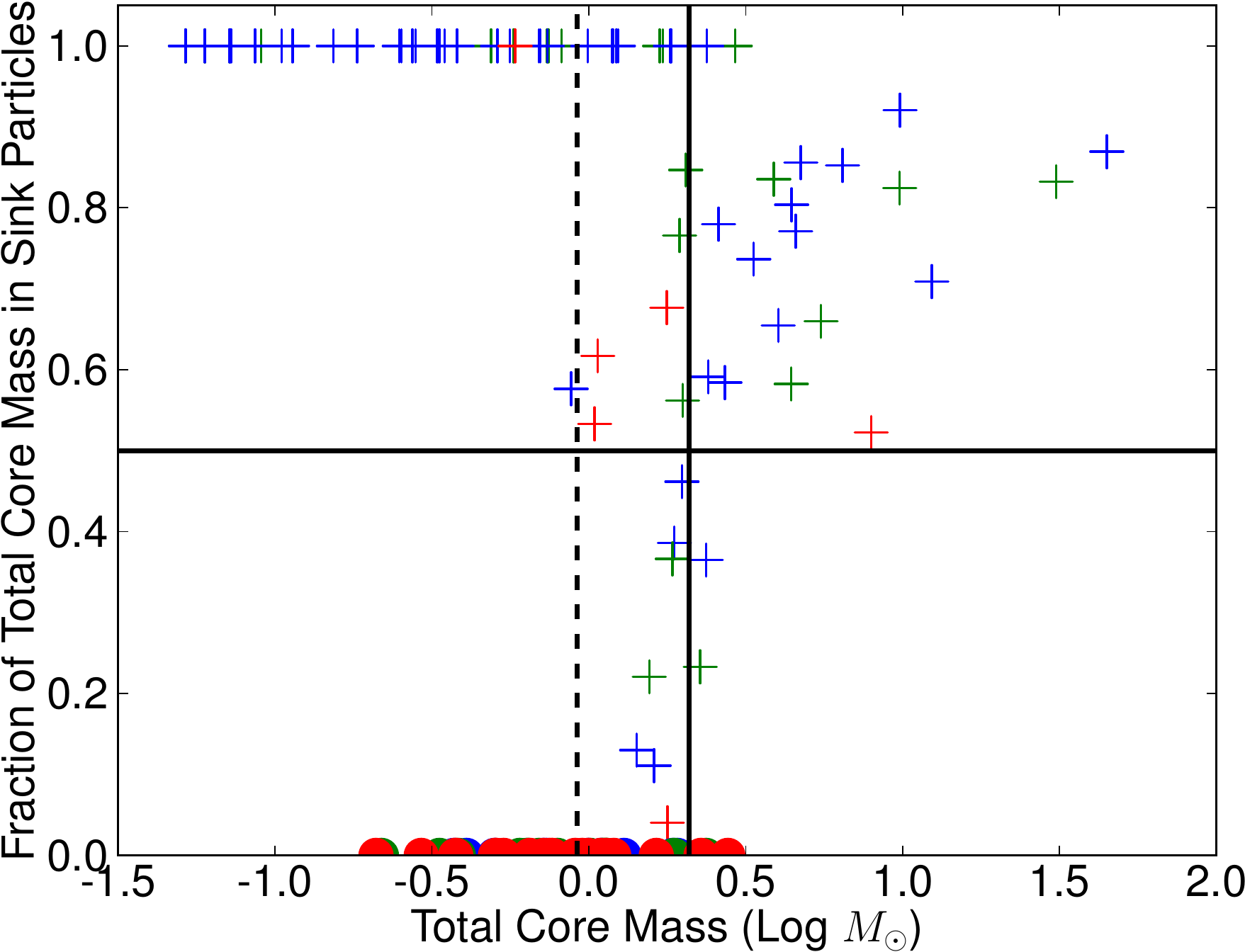}
\caption{The fraction of a total core's mass (protostar and envelope) found in the protostars contained within the object's boundaries plotted against total core mass for all objects in all three projections observed at three timesteps. The Y dimension integrated images only are shown here for clarity. Red represents 50$\%$ of the box free-fall time, green represents 80$\%$ of the box free-fall time, and blue represents one free-fall time. The solid vertical line is drawn at the Jeans mass corresponding to the shocked density (see section \ref{densities}). The dashed vertical line highlights the Jeans mass associated with the empirically derived ``modal density''. The solid horizontal line simply shows the 50$\%$ mark (i.e. where the collapsing regions dominate the core mass).}
\label{sinkmassfrac}
\end{figure}

				 In this section, we explore the relationship between the mass of a protostar and its parent core. Figure \ref{sinkmassfrac} shows the fraction of core mass in sink particles. The points which lie at exactly 1.0 on the ordinate axis are protostars which did not have an associated envelope. These protostars lie outside 75$\%$ of their nearest core's radius (measured from the core centre). Thus, with no apparent envelope, the sink mass is one hundred percent of the object's mass. Note that because of the background subtraction, some of the core mass may be lost in this analysis.

                                  As expected, many young objects have not formed protostars and lie along the bottom of the plot. As time progresses,
				the sink particles begin to dominate the core masses quickly. The bottom right quadrant of Figure \ref{sinkmassfrac} is empty. 
				This means that the more massive cores observed in the simulation are dominated by the protostars present. Stepping
				through time, it is clear that this is a rapid process. Once sink particles form, they quickly accrete a large amount of mass. Therefore,
				observations ignoring the embedded protostars could miss a significant portion of mass. This will
				lead to errors in stability classification using non-interferometric observational techniques. 
				
				When each of the timesteps are analysed in a sequential fashion, the points travel from the bottom left through an ``S'' shape to the top right of the plot.
				Since the total mass of a given object substantially increases throughout the formation and growth of a protostar, it is clear that the protostar mass cannot come from the initial material detected in the envelope alone. One possibility is that cores extend below our flux limit and are collapsing to higher densities such that they enter the observable regime. Another likely possibility is the accretion of mass due to bulk flows in the simulation in conjunction with gravity. The obvious filamentary structures in the map are the most likely sources of mass. In fact, mass flow along filaments associated with protostars have been observed in many star forming regions (e.g. \citealt{kirkfilamentflow, friesenfilamentflow, hacarfilamentflow, schneiderfilamentflow}).

			Since the final mass of the protostars is sometimes larger than the initial observed core mass, this suggests either observed cores are initially more massive than observed or that they continue to accrete. In either case, this undermines the comparison of a single time snapshot of the core mass distribution with the IMF.
				
\vspace{0.3cm}

\section{Interferometric Analysis}
\label{ALMAsec}

In this section, we assess the conditions for which the detection of an embedded protostar is possible. Employing the high resolution and sensitivity offered by interferometers is the logical next step in characterising the dynamic nature of cores. 

\begin{table*}
\caption{ALMA Cycle 1 observations performed on three cores.}
\centering
\begin{tabular*}{0.9\textwidth}{@{\extracolsep{\fill}}  c  c  c  c  c  c }
\hline\hline
Core Identifier & Radius (AU) & Envelope Mass ($M_{\rm \odot}$) & Sink Particle Mass ($M_{\rm \odot}$) & $\frac{\mathrm{Envelope\:Mass}}{M_{\rm J}}$ & Density (cm$^{-3}$)\\\hline
1 & 829.30  &  0.27  & 0.619 & 2.44  & 1.70 x 10$^{7}$\\
2 & 579.48  &  0.09  & 0.090 & 1.17 & 1.66 x 10$^{7}$\\
3 &  645.38 &  0.13  & 0.078 & 1.57 & 1.81 x 10$^{7}$\\\hline
\end{tabular*}
\label{almadetection}
\end{table*} 

As described in section \ref{interferometermethods}, we performed synthetic ALMA Cycle 1, SMA, and CARMA simulated interferometric observations. It was found, however, that a ninety second observation performed at 100 GHz with ALMA Cycle 1 is comparable to an eight hour observation taken by SMA at 230 GHz and achieves a far better signal to noise ratio than an eight hour observation taken with CARMA at 100 GHz. \cite{offnerinterferometers} found a similar result. The SMA and CARMA observations produced only non detections; evidently, to identify any substructure present in faint cores or to even detect the objects with $3\sigma$ confidence, a greater sensitivity is required. This null result is compatible with recent interferometric observations (see \citealt{lackofsubstructureschnee}).

Thus, we focus primarily on ALMA. We analyse each interferometric image down to the same flux threshold ($6\sigma=0.6$ mJy/beam, defined by-eye) for a consistent analysis. Purely investigating the results of CLFIND2D, we see that in most cases there is no significant buildup of mass before a protostar forms. In two of the cases the core is identified very briefly after the formation of the protostar. In one case, the core is identified concurrently with the formation of a sink particle and continues to grow in time. In the last case, CLFIND2D does not identify a core at any time output.
 
\begin{figure*}
\centering
\includegraphics[]{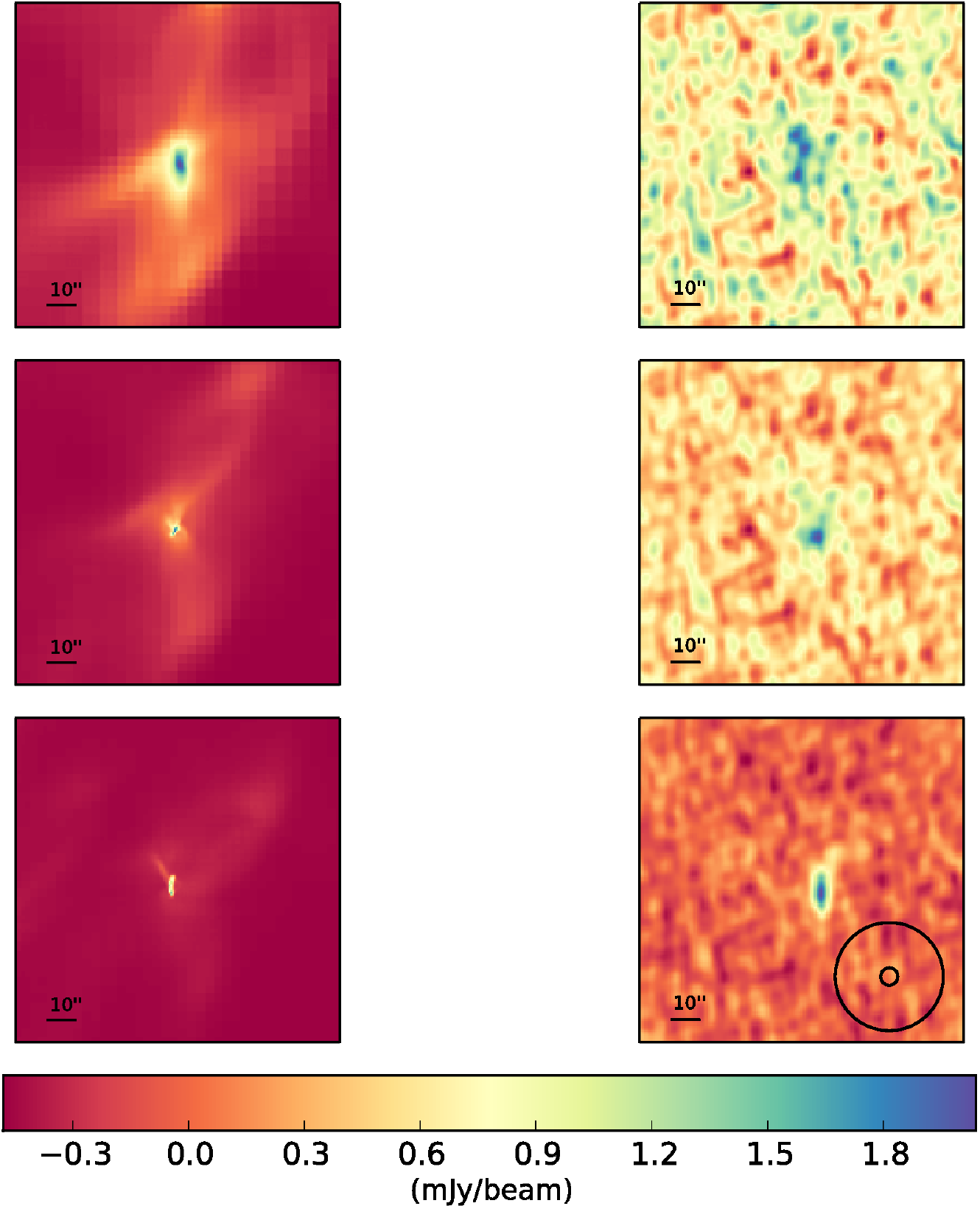}
\caption{ALMA Cycle 1 simulated observations of Core 1. The left column shows three original, simulated, images at different timesteps. The right column shows the interferometric observations of these same three timesteps. The top and middle rows show times 0.20$t_{\rm ff}$ and 0.24$t_{\rm ff}$. The third time shows the object at 0.29 $t_{\rm ff}$, just after a protostar has formed. The large circle on the bottom right hand panel represents the effective 20\arcsec $\:$smoothed beam in the single dish analysis. The smaller circle shows the 3.2\arcsec $\:$100 GHz synthesised ALMA beam.}
\label{ALMAfullycore0}
\end{figure*}

 Figure \ref{ALMAfullycore0} shows the observations of the object in the bottom left panel of Figure \ref{random} (labeled ``Core 1'' in Figure \ref{simfig}) at three different timesteps; two approaching the formation of the first protostar and one at the time the first protostar appears. The core begins in an undetected state. By the time a protostar forms, a clear detection is possible. 

Table \ref{almadetection} shows the properties of the three identified objects at the point of their first detection. In all three cases, the objects already contained sink particles. Cores 1 and 2 are the same as in Figure \ref{random} (left and right panels, respectively); note the increase in density in the centre of the core compared to the core average (compare with Section \ref{densitytrackssection}).        

 CASA simulations such as these provide a strong prediction for real interferometric observations. Currently, ALMA Cycle 1 telescope time has been awarded with highest priority to observe the 3mm continuum emission from all 60 starless cores and 13 protostellar cores in the Chamaeleon I molecular cloud, as identified in \cite{belloche2011}. These observations will be sensitive to point sources with masses $\gtrsim 0.01 \mathrm{\:}M_{\rm \odot}$, with less than 2 minutes of on-source integration time per object (see Table \ref{almadetection}). Once these data are collected, we will be able to perform robust comparisons between simulated core properties and their observed counterparts.

\section{Discussion}
\label{discussionsec}

\subsection{Shocked Densities and Structure}
\label{densities}

In this section we consider some simple estimates of the role of turbulence and gravity to place our results in context. 

The characteristic size scale at which gravity dominates over thermal pressure is given by: 

\begin{equation}				
				L_{\rm J} = \sqrt{\frac{\pi c_{\rm s}^{2}}{G\rho_{\rm 0}}}.
\end{equation}
where $c_s=\left(k_{\rm B}T/\mu m_{\rm H}\right)^{\frac{1}{2}}$ is the sound speed, $\mu=2.33$ is the mean molecular weight, $k_{\rm B}= 1.38$ x $10^{-16}$ ergK$^{-1}$, G is the gravitational constant,  $m_{\rm H}$ is the mass of hydrogen, T is the isothermal gas temperature, and $\rho_{\rm 0}$ is the average mass density in the simulation.  
				
				
				In a typical shocked region in the simulation, the density and hence the Jeans length will be higher. The shocked density of a 1D isothermal shock is given by $\rho_{\rm s} = \mathcal{M}_{\rm 1D}^{2}\rho_{\rm 0}$, where $\mathcal{M}_{\rm 1D}=\mathcal{M}_{\rm 3D}/\sqrt{3}=3.8$ is the 1D simulation Mach number.  Plugging this into Equation 3, we can expresses the  compressed Jeans length as:

\begin{equation}
                                L_{\rm Jm} = \sqrt{\frac{\pi c_{\rm s}^{2}}{G\mathcal{M}_{\rm 1D}^{2}\rho_{\rm 0}}} 
\end{equation}	
 				or,
					
\begin{equation}				
			      L_{\rm Jm} = \mathcal{M}_{\rm 1D}^{-1}L_{\rm J}. 
\end{equation}

				Values of $T = 10$ K and $\rho_{\rm 0} = 5.1 \mathrm{\:x\:} 10^{-21} \mathrm{\:g\:cm}^{-3}$ give a compressed Jeans length of $L_{\rm Jm} = 4.7 \mathrm{\:x\:} 10^{17}$ cm or 126.42\arcsec, assuming a distance of 250 pc. For the 2 pc box, the low resolution (512 x 512) grid used in this analysis had 3.2 x 3.2 arcsecond pixels. So, a typical shock forces material together on scales of 
				$\sim$40 pixels in the low resolution grid. 
				
				The Jeans mass, $M_{\rm J}$, associated with the Jeans length is given by Equation \ref{jeansmasseq} replacing $R_{\rm c}$ with the Jeans radius, $R_{\rm J} = L_{\rm J}/2$. Thus, the Jeans mass corresponding to the average density in the simulation is $M_{\rm J,0} = 7.92 \mathrm{\:}M_{\rm \odot}$ and the Jeans mass corresponding to shocked densities is $M_{\rm J,m}=2.08 \mathrm{\:}M_{\rm \odot}$. The average CLFIND2D core mass is much closer to the shocked Jeans mass than the Jeans mass corresponding to the average density of the box. 

\begin{figure*}
\centering
\includegraphics[width=16cm,height=16cm]{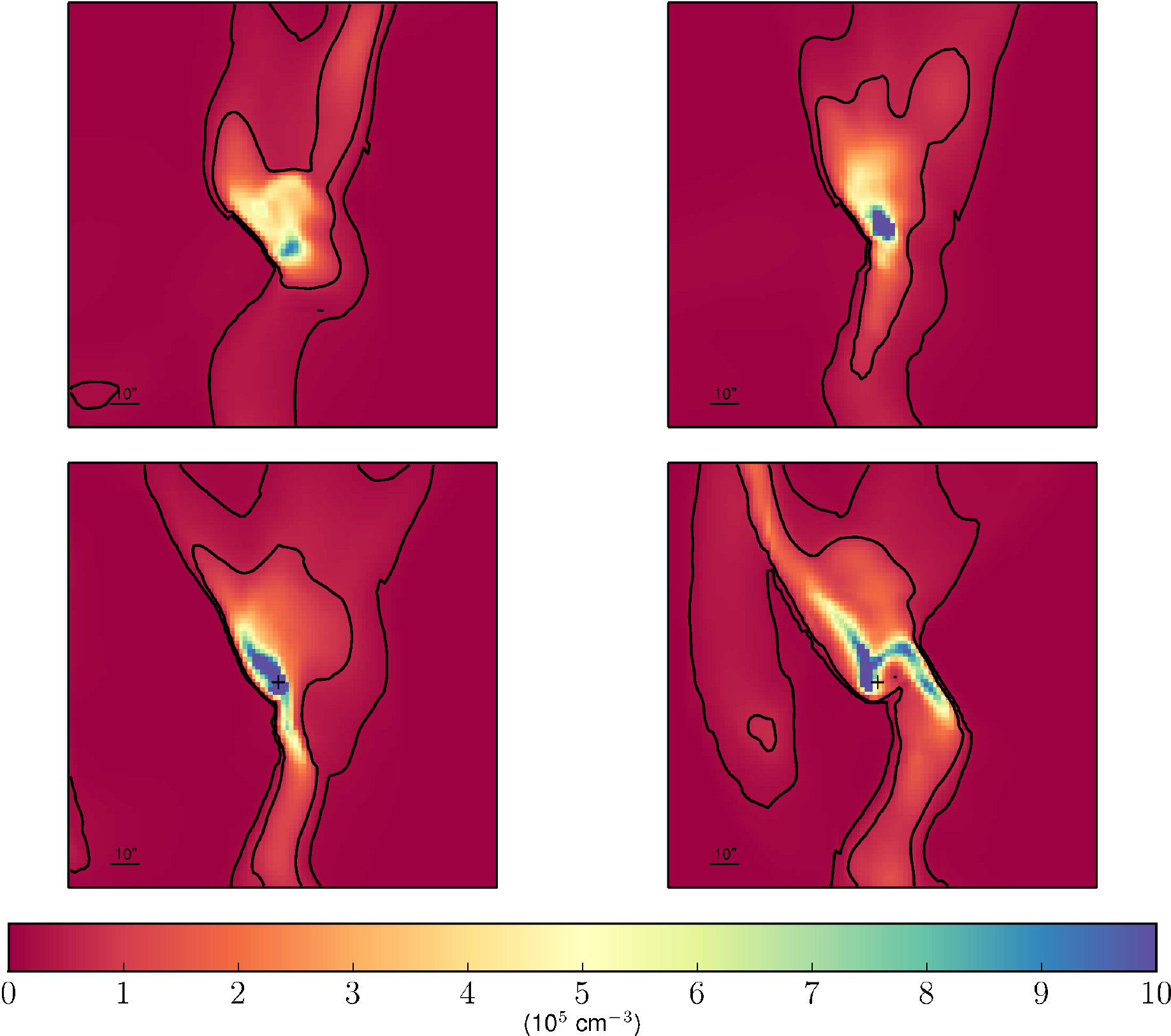}
\caption{Density map for one individual core observed at a resolution of 1.6$\arcsec$/pixel. The top row shows the core before it forms a protostar (Left to right: 0.15$t_{\rm ff}$ and 0.24$t_{\rm ff}$); the bottom row shows two timesteps shortly after a protostar forms (Left to right: 0.34$t_{\rm ff}$ and 0.41$t_{\rm ff}$). The plus signs indicate the locations of the protostar. The outer contour indicates a density of $1.9 \mathrm{\:x\:} 10^{4}$ cm$^{-3}$; the inner: $1.0 \mathrm{\:x\:} 10^{5}$ cm$^{-3}$ (see text).}
\label{densityfig}
\end{figure*}				
							
				This is, of course, a very simple estimate. The actual processes at play are more complicated and can involve a number of oblique shocks and shears. 
								Note that the ``modal density'' is approximately a factor of 5 greater than the shocked density calculated here; this corresponds to a higher characteristic velocity 
								 than that associated with $\mathcal{M}_{\rm 1D}$.    
				
				 To gain a better understanding of the true densities attained in the centre
				of an individual object,  we analysed one of the cores tracked in Sections 4.3 and 4.4 in more depth at a higher resolution (level 2 in the AMR simulation). The core we selected was the first object to achieve a super-Jeans state in the Y-dimension flux maps.
				
				Figure \ref{densityfig} shows number density maps for Core 1 at four different timesteps centred on the formation of the first sink. The outer contours show the expected number density achieved using the $\mathcal{M}_{\rm 1D}^{2}$ coefficient ($n_{\rm H} = 1.9 \mathrm{\:x\:} 10^{4} \mathrm{\:cm}^{-3}$). The inner contours show the ``modal density'' ($n_{\rm H} = 1.0 \mathrm{\:x\:} 10^{5} \mathrm{\:cm}^{-3}$); approximately a factor of 5 larger. Clearly, both these values are significant in tracing the structure present in the map.
							
				 Evidently, much higher densities are achieved than those expected by the simple argument presented above. Throughout the observed frames, both before and
				after the protostar has formed within the core, gravity quickly creates regions of high density on scales much smaller than the beam. Thus, important structure will remain unobserved without deep interferometric maps. 
				
				Analysing the simulation itself, a typical number density value within Figure \ref{densityfig}'s inner contour was found to be $n_{\rm H} = 5 \mathrm{\:x\:} 10^{5} \mathrm{\:cm}^{-3}$ by averaging over the projection. The highest inferred average density of a region exceeds $n_{\rm H} = 10^{6} \mathrm{\:cm}^{-3}$ (see Table \ref{almadetection}). This typical density corresponds to a dynamical collapse time of approximately 5\% of the total box free-fall time. Near the beginning of the simulation, this is close to the time resolution in which observations were performed. Therefore, for densities significantly greater than  5 x 10$^{5}$ cm$^{-3}$, the collapse will not be resolved temporally with the chosen, observed simulation snapshots in the earlier stages of core evolution.
				
\subsection{Single Dish Results}
\label{singledishresults}

The CLFIND2D objects span a range of masses and morphologies. The objects are initially detected when they are starless; quickly, a subset undergoes mass accretion, gravitational collapse, and protostar formation. In fact, the majority of the cores detected go on to form protostars. This indicates that the objects observed are not transient, but ``real'' star formation sites. In the earliest timesteps, before gravity has had a chance to significantly affect localised regions, there are very few cores identified. As time progresses, the number of observationally defined super-Jeans cores as well as the number of protostars increases.

                                It is interesting to note that the majority of the mass in cores accretes onto the protostars and the protostar mass rapidly dominates the mass budget. In fact, in this simulation the most massive envelope including at least one protostar was approximately 10 $M_{\rm \odot}$. The cumulative mass of the protostars present within the largest core, however, totalled more than $60 \mathrm{\:}M_{\rm \odot}$. Recall, however, that without removing the 120\arcsec $\:$scale structure from the flux maps, we find these cores reside within much larger mass envelopes (Section \ref{clumpfind}). These extended regions predominately trace the filamentary structure in the simulation and act as reservoirs for the smaller scale cores.
                                
                                When the mass of the protostars is not included (to resemble observations of cores wherein the dense central object is unobservable) there exists a small population of cores forming stars which are deemed stable from an observational perspective. When the protostellar mass is added, however, these cores are found to be unstable to collapse as expected.
                                
                                Observationally detecting collapse proves to be quite difficult. For example, if the protostar is deeply embedded and undetectable, its mass cannot be accounted for in the gravitational analysis. The problem of detecting embedded protostars is generally expected to be more severe when the protostars are small and dim. Consequently, the results presented here are most applicable to the transition between the starless and protostellar stage. In many cases, the core, the protostar, or both may be too faint to detect in the first place; and detection is especially difficult at early times \citep{lackofsubstructureschnee,pineda2011,dunham2008,bourke2006,young2004}.
                               
                                 \cite{smith2012} show that the problem of detecting protostars extends to observations of molecular lines. They performed radiative transfer calculations for cores embedded in filaments in a turbulent hydrodynamic simulation. They find that in over 50\% of viewing angles, there is no ``blue asymmetry'': a classic sign of material infall in an isolated spherical core. In a continuation of this work, \cite{smith2013} highlight the need for high resolution observations with ALMA in order to test how line profiles and results change with beam size.
                                
                                Note that a few significantly super-Jeans starless objects (core mass $> 4.5M_{J,c}$) were identified by \cite{sadavoy2010}, little evidence of objects fitting this classification is found in this simulation.
                                
                                                                Of course, the detection and analysis of cores relies on many assumptions. We adopted typical values for the dust properties \citep{dougoph,kirkperseus,howstarlessschnee,sadavoy2010big}.  We neglected internal heating due to protostars. In actual observations, the warmed dust grains cause an increase in flux which can easily be misconstrued as a core with a larger mass. 
                                
                                One can clearly see a low mass core population which is maintained throughout the simulation in Figure \ref{hists}. It appears that once a core is identified, it continues to collapse and gain mass. Most of this mass is accreted onto embedded protostars. A sufficient amount of the mass flow, however, replenishes the dust envelope, leaving it detectable. 
                                
                                This distribution of mass within an evolving protostellar core can be best illustrated by Figures \ref{hists} and \ref{sinkmassfrac}. The top right hand panel of Figure \ref{hists}, which takes into account only what is observable in the submillimetre regime, shows the total remains approximately constant over time. If the mass of protostar is included, the total mass within the core increases over time. Even when the embedded protostars become quite massive, the core masses remain comparatively low.  When only considering the material in the envelope, the density increases only slightly over time. In general, when the protostellar masses are added, the peak of the mass and density distributions stay approximately constant while more higher mass objects are observed. This is also shown in Figure \ref{sinkmassfrac} where the protostars dominate the mass by a significant factor.
                                
                                 These distributions were also analysed assuming each protostar was a factor of 3 less massive in order to compensate for the overestimate inherent in the simulation (see Section \ref{datasec}). We found a less accentuated but still significant increase in higher mass objects as expected. A modified version of Figure \ref{sinkmassfrac} did not undergo any substantial changes. The protostar masses still greatly dominate the envelope masses.
 
 \vspace{0.5cm}
 
 \subsection{Interferometric Results}                            
\label{interferometers}

To truly determine the nature of a given core, one needs to look more closely at the internal structure. The large JCMT beamsize ``washes out'' the more compact structure, causing the object to appear less dense. In order to investigate the details of protostar formation at the scales, sensitivity, image fidelity, and resolution necessary, an interferometer such as ALMA is required.
				
				The bottom right panel of Figure \ref{ALMAfullycore0} shows the difference between the JCMT and the ALMA beams for a simulated ALMA Cycle 1 observation (100 GHz). It is clear that the JCMT beam blends much smaller structures that {\it can} be detected with an interferometer. 

				Comparing the Table \ref{almadetection} values with the same cores as in the single dish data, we highlight several interesting points. Beginning with Core 1, the mass inferred using interferometry is an order of magnitude less than found by the JCMT. This is to be expected as the envelope observed by ALMA is much smaller than that observed by the JCMT. The reduction in mass indicates that there is structure present that is on larger angular scales that cannot be recovered by ALMA's 12m array in its most compact configuration at 100 GHz. 
				The object was detected concurrently with the snapshot in which the protostar formed.  
				
				Core 2 appears more unstable when synthetically observed by ALMA. Removing the noise from the observations and redefining the clump boundaries at a lower flux threshold which is discernible by-eye, the object is found to be up to 1.5 times more massive. This indicates that the signal to noise ratio significantly influences the core
classification and stability calculation. 
The object was detected shortly after the sink particle first appeared. The third isolated core in which a detection was made with ALMA is similar to the second.	 The core mass to Jeans mass ratio indicates more instability in the case of ALMA. Probing the central densities of these cores is important for understanding the single dish data. 
		
		Our interferometric analysis of core stability also assumes isothermal dust and spherical cores. It is clear, however, that the morphologies of these objects could indicate collapse along preferred axes (see \citealt{pon2011} for a thorough description of collapse modes). This type of analysis is vital to perform at small scales near the centre of identified objects in order to truly observe how stars are forming. 
 
\section{Conclusion}	
\label{conclusionsec}

In this study, we performed synthetic single dish and interferometric observations of a simulated star forming region.  Assuming the gas and dust were optically thin, we inferred masses and densities by assuming the objects were spheres. We calculated the gravitational stability using the density of each core and correlated protostars with the objects. The single dish analysis was performed with and without including protostellar masses; the former was to emulate real observations while the latter was to compute the core's ``true'' stability. We investigated the relationship between the protostar and its parent envelope in terms of their mass and we considered the significance of the core densities with respect to turbulence and gravity.

There are several key results:

\begin{enumerate}[start=1]

\item Our analysis is consistent with various observations. Namely, the masses and densities of the simulated cores we detect are very similar to ``real'' cores in Perseus (see Section \ref{bulk}). We find many more sub-Jeans cores than super-Jeans, in terms of their mass and size, which is consistent with \cite{sadavoy2010big}. The fact that we do not detect substructure with simulated CARMA observations is consistent with \cite{lackofsubstructureschnee}.  

\item Nearly all cores that we detect eventually form protostars. This suggests that observed cores detected in this manner (assuming the physical conditions and distance of Perseus) are probably ``real''; that is, they will likely go on to form protostars in the future (see Section \ref{singledishresults}). The mass of the observed envelope, however, does not appear to be a good tracer of the eventual protostellar mass. This has implications for comparing the core mass function with the stellar initial mass function (see section \ref{sinkenvelope}). Note, however, that the simulation does not include magnetic fields, which could provide additional support that may inhibit a core from collapsing.

\item Nearly all cores we identify are associated with filaments. This is consistent with the ubiquity of filaments recently observed by Herschel and suggests that if observations had better resolution and sensitivity they would also see a similar correspondence between cores and filaments (see Section \ref{filaments}). 

\item Single dish observations such as those with the JCMT as well as previous-generation interferometers appear to miss significant core structure on small scales due to flux averaging. Interferometric observations with ALMA are necessary to recover this information (see Sections \ref{ALMAsec} and \ref{interferometers}).

\end{enumerate}

\section{Acknowledgements} 

Doug Johnstone is supported by the National Research Council of Canada and by a Natural Sciences and Engineering Research Council of Canada (NSERC) Discovery Grant. Support for this work was provided by NASA though Hubble Fellowship grant \#51311.01 awarded by the Space Telescope Science Institute,  which is operated by the Association of Universities for Research in Astronomy, INC, for NASA, under contract NAS 5-26555 (SSRO). The simulation was performed on the Trestles XSEDE resource. The National Radio Astronomy Observatory is a facility of the National Science Foundation operated under cooperative agreement by Associated Universities, Inc.

\bibliography{starless.bib}

\end{document}